\documentclass[12pt]{article}
\usepackage{kpfonts}

\usepackage[T1]{fontenc}
\usepackage{sfmath}
\usepackage{helvet}
\usepackage{amsmath}
\usepackage{amsthm}
\usepackage{enumitem}
\usepackage{todonotes}

\usepackage{amssymb}
\usepackage{geometry}
\usepackage{soul}
\usepackage{xspace}
\newtheorem{claim}{Claim}
         
\newtheorem{proposition}{Proposition}       
\newtheorem{remark}{Remark}  
\newtheorem{example}{Example}  
\newtheorem{theorem}{Theorem}  
\newtheorem{procedure}{Procedure}  
\newtheorem{lemma}{Lemma}  
\usepackage{hyperref}
\usepackage{xcolor}
\hypersetup{
    colorlinks,
    linkcolor={orange!75!black},
    citecolor={violet!90!black},
    urlcolor={blue!80!black}
}

\usepackage{comment}

\usepackage{verbatim}
\usepackage{appendix}

\usepackage{natbib}
\newcommand{\Csub}{\mathcal C^{\text{sub}}}
\newcommand{\Cpi}{\mathcal C^{\text{pi}}}
\newcommand{\Cres}{\mathcal C^{\text{res}}}
\newcommand{\Csvres}{\mathcal C^{\text{sv}-\text{res}}}
\newcommand{\Ccon}{\mathcal C^{\text{con}}}

\newcommand{\Csubsm}{\mathcal C^{\text{sub-sm}}}
\newcommand{\Chres}{\mathcal C^{\text{mto1-res}}}

\newcommand{\Chpi}{\overline {\mathcal C}^{\text{pi}}}
\newcommand{\Chsubsm}{\overline {\mathcal C}^{\text{sub-sm}}}
\newcommand{\Ch}{\mathcal C^{mto1}}

\newcommand{\mto}{many-to-one\xspace}
\newcommand{\exclusion}{exclusion\xspace}
\newcommand{\IRC}{consistency\xspace}
\newcommand{\Ens}{\mathcal E}

\makeatletter
\renewcommand\paragraph{\@startsection{paragraph}{4}{\z@}%
                                    {2ex \@plus1ex \@minus.2ex}%
                                    {-1em}%
                                    {\normalfont\normalsize\bfseries}}
\makeatother

\title{Lexicographic Composition of Choice Functions}
\author{
\begin{tabular}{c}
  Sean Horan\\
{\footnotesize Universit\'e de Montr\'eal}\\
{\footnotesize sean.horan@umontreal.ca}
\end{tabular}
\hspace{1in}
\begin{tabular}{c}
Vikram Manjunath\\
{\footnotesize University of Ottawa}\\
{\footnotesize vikram@dosamobile.com}
\end{tabular}}

\usepackage{cleveref}
\crefname{claim}{claim}{claims}
\Crefname{claim}{Claim}{Claims}
\crefname{procedure}{procedure}{procedures}
\Crefname{procedure}{Procedure}{Procedures}

\begin{document}

\maketitle
\begin{abstract}

Lexicographic composition is a natural way to build an aggregate
choice function from component choice functions. As the name suggests,
the components are ordered and choose sequentially. The sets that
subsequent components select from are constrained by the choices made
by earlier choice functions. The specific  constraints affect whether
properties like path independence  are preserved. For several domains
of inputs, we characterize the constraints that ensure such preservation.

\end{abstract}
\section{Introduction}

Given a set of options, a choice function selects a subset. Choice
functions are one of the elementary building blocks of microeconomic
theory. They are a way to represent  individual preferences or even
aggregated social preferences and priorities.\footnote{See, for
  instance, \cite{MoulinSCW1985}} In contexts with competing preferences and
priorities, choice functions need to be aggregated and a natural way
to do so is \emph{lexicographically}.
That is, given a list of component choice functions, the first one selects from
the full set of options, the second one selects from what is \emph{feasible}
given the first choice, and so on. The final choice of the aggregated
choice function is
the union of what each individual choice function selects. We focus on
feasibility,
in terms of constraints that earlier choices place on later
choices. So, we frame feasibility in terms of  
\emph{exclusions}. Depending on what has previously been chosen,
certain things are excluded from what can be chosen  at a particular step
of the lexicographic process. Specifically, an exclusion function $E$
maps every set of items to another set of items. If a set $Z$ of items
has previously been chosen, then $E(Z)$ is excluded from the input to
the next choice function. That is, if we start from $Y$ and $Z$ has
already been chosen, then the next choice function gets to choose from
$Y\setminus E(Z)$.
Here are some examples:
\begin{itemize}
\item Indivisible private goods: $E$ is the identity
function.
\item Maximizer-collectors \`a la
\cite{AizermanMalishevski:IEEE1981}: $E$ always maps to the empty
set.
\item Contracts as in \cite{RothECMA1984} and
  \cite{HatfieldMilgrom:AER2005}: 
$E$ maps to the set of contracts with ``doctors'' named in previously
chosen contracts.
\item Capacity constraints: $E$ is identity if fewer items have been
  chosen than the capacity or the universal set otherwise.
\end{itemize}

Whatever the feasibility constraints are, lexicographic composition is
equivalent to \emph{serial   dictatorship} 
over the competing interest. If these interests are
strategic, then concerns about incentive compatibility and efficiency often
narrow the possibilities  for aggregation to lexicographic composition
  \citep{Svensson:1999SCW,
    Papai:JPET2001,EhlersKlaus:SCW2003,Hatfield:SCW2009,PyciaUnver:2022}. 
  Even without strategic considerations, such composition is a common
  tool in the design of choice functions  in the market design
  literature.\footnote{See, 
    \cite{SonmezSwitzer:2013ECMA}, \cite{Sonmez:JPE2013}, and
    \cite{KominersSonmezTE2016} among others.} In fact, this
  hierarchical structure is, in a sense, 
necessary for organizational choice functions to be such that stable
matchings are guaranteed to exist
\citep{Alva:2016}.

Properties like path independence have
long been studied in choice theory as a weak form of rationality
\citep{PlottEconometrica1973}. Path 
independence is specifically 
 important in matching settings as pointed out by
\cite{ChambersYenmezAEJMicro2017}.\footnote{Path independence is
  equivalent to the combination of \emph{substitutes} and \emph{consistency} (also
  called  ``independence of rejected contracts'' \citep{AygunSonmez:AER2013})
  \citep{AizermanMalishevski:IEEE1981}.}
Our interest is in understanding what exclusion functions have the
property that the lexicographic   composition of path independent choice
functions is itself path 
independent. It is well known that for the first example above, of the
identity exclusion, path independence is preserved by lexicographic
composition \citep{Alva:2016,ChambersYenmezAEJMicro2017}. At the other
extreme, \cite{AizermanMalishevski:IEEE1981} show that every path
independent choice function can 
be rationalized by the lexicographic composition of single-valued
rational choice functions with the \emph{empty} exclusion
function.\footnote{The exclusion function that maps every set to the
  empty set.}  However,
not all exclusion functions have this property. Famously, exclusion
based on equivalence relations---like the
``contracts'' exclusion described above---does not
\citep{HatfieldMilgrom:AER2005,HatfieldKojima:JET2010}.

Our main contribution is to characterize the exclusion functions such
that the lexicographic composition of two responsive choice functions
is path independent (\Cref{thm: cres}).\footnote{A choice function is
  responsive if there 
is an ordering $\succ$ over the items and a capacity such that
from any set of items, it selects the best items according to $\succ$,
up to its capacity. Every responsive choice function is path
independent. So, the necessary conditions from this result extend to
larger domains than responsive choice functions.} 
We then consider two variants of this result. The first is to expand
the domain from responsive choice functions to arbitrary path
independent choice functions (\Cref{prop: pi}). The second is to
consider the preservation of an additional property, size
monotonicity, which is commonly studied in the matching
literature (\Cref{prop: sm}).\footnote{Size   monotonicity is
  sometimes  referred to as the  ``law of aggregate demand''
  \citep{HatfieldMilgrom:AER2005}.}  In both cases, the set of
surviving exclusion functions shrinks.

While our main results are stated for the composition of just two choice
functions, nested applications allow one to aggregate more of
them. In general, the binary operation of lexicographic composition is
not associative or commutative. Consequently, there is a great
richness to the ways a list of $n$ choice functions can be aggregated:
they can be mapped to binary trees with $n$ leaves. Moreover, each
non-terminal node in such a tree can be labeled by an exclusion
function.

The aggregation of single-valued, rational choice functions is
particularly salient \citep{KominersSonmezTE2016}. So, we also characterize
the exclusion functions that preserve path independence when one of
the two inputs is single-valued (\Cref{prop:sv-sub} and
\Cref{prop:sub-sv}).

As noted above, exclusion functions based on equivalence relations
violate the 
necessary conditions to preserve path independence. So, we consider
the weaker property of path independent completability
\citep{HatKom:2014HSubs}. We show that the obvious adaptations of the
conditions for preserving path independence are sufficient
 to preserve path independent completability (an application of
\Cref{lem:mto1}).

The remainder of the paper is organized as follows. In
\Cref{sec:definitions} we set up the model. In \Cref{sec:path
  independence} we give our main result.  In 
\Cref{sec:addit-prop-broad} we examine the consequences of expanding
the domain and of preserving size monotonicity along with path
independence. We consider the composition of more than two choice
functions in \Cref{sec:nested}. In \Cref{sec:equiv}, we consider equivalence relation based
exclusion. We end with a brief discussion in  \Cref{sec:discussion}. All  proofs are in
the appendix.

\section{Definitions}
\label{sec:definitions}

Let $X$ be a countably infinite set of
\textbf{items}. Let $[X]$ be the power set of $X$ and
$[X]^*=\{ Y \in [X]: |Y| < \infty \}$ be the set of finite subsets of
$X$.

A \textbf{set function} $S: [X]^* \rightarrow [X]$ is a mapping
from finite subsets of $X$ to subsets of $X$. Let $\mathcal S$ be the set of
all set functions. 

Two kinds of set function are particularly relevant for our analysis.
First, a \textbf{dilation} $D \in \mathcal S$ is
a set function such that, for each $Y \in [X]^*$, $D(Y) \supseteq
Y$.\footnote{\cite{DanilovKoshevoyOrder2009} call  these
  ``extensive operators''. } Conversely, a \textbf{choice function} (or
\textbf{contraction}) $C \in \mathcal S$ is 
a set function such that, for each $Y \in [X]^*$, $C(Y) \subseteq
Y$. Let $\mathcal D$ and $\mathcal C$ be the set of all dilations and choice
functions, respectively.

The \textbf{lexicographic composition} $\mathcal L_{E}(C_1,C_2) \in
\mathcal C$ of two choice functions $C_1,C_2 \in \mathcal C$ subject
to the set function $E\in \mathcal S$ is defined, for each $Y
\in [X]^*$, by 
\[
  \mathcal L_{E}(C_1,C_2)(Y) =C_1(Y) \cup C_2(Y\setminus E(C_1(Y))).
\]
The first choice function $C_1$ chooses from all of $Y$. Given the
alternatives $C_1(Y)$ that it chooses, the set function $E$ specifies
the alternatives $E(C_1(Y))$ to be 
excluded from being chosen by the second choice function 
$C_2$. Because of the role that it plays in the definition of
$\mathcal L_{E}$, we refer to $E$ as an \textbf{\exclusion
  function}. 

Our  interest is in understanding for which  $E$s the lexicographic
composition $\mathcal L_{E}$  inherits certain properties of the 
input choice functions $C_1$ and $C_2$.
For a particular property $\pi$, let $\mathcal C^\pi \subseteq
\mathcal C$ denote the set of choice functions that satisfy
$\pi$. Given a pair of properties, $\pi$ and $\rho$, we denote the conjunction
 by $\pi\text{-}\rho$ so that 
$C^{\pi\text{-}\rho}$ is the set of  choice functions that satisfy both $\pi$ and
$\rho$.\footnote{That is, $C^{\pi\text{-}\rho} =
  C^{\pi}\cap C^{\rho}$.}
Then, given a set of choice functions 
$\mathcal C' \subseteq \mathcal C^\pi$, the lexicographic composition
\textbf{\boldmath $\mathcal L_{E}$ preserves $\pi$ over $\mathcal C'$} if
$\mathcal L_{E}(C_1,C_2) \in \mathcal C^\pi$ for all $C_1,C_2  \in
\mathcal C'$. In some instances, we consider cases where the
  two inputs are from different sets. In such contexts, we speak of
  $\mathcal L_E$ preserving $\pi$ over $\mathcal C_1\times \mathcal
  C_2$ for some $\mathcal C_1, \mathcal C_2\subseteq \mathcal C^\pi$.
  
We do not consider the preservation of a property $\pi$
  outside of $\mathcal C^\pi$ since the composition cannot satisfy
  $\pi$ if the inputs are not ensured to. To see why, suppose that
  $C_2$ always chooses the empty set. Then the composition of any
  $C_1$ with $C_2$ is just $C_1$, no matter what $E$ is. If $C_1$
  violates $\pi$, then the composition does as well. While we could
  impose joint conditions on $C_1$ and $C_2$ to ensure that their
  composition satisfies $\pi$, we feel that it is beyond the scope of
  this paper.

\section{Path Independence and Responsive Choice}
\label{sec:path independence}
 A choice function $C\in \mathcal C$ is \textbf{path independent (pi)}
 if for any set of items $Y\in [X]^*$, the choice from $Y$ is
 invariant to arbitrarily segmenting $Y$ into
 several parts, choosing from each part, and then choosing again from all of
 the chosen items. That is,  for each pair $Y,
 Y'\in [X]^*$, $C(Y\cup Y') = C(C(Y) \cup C(Y'))$
 \citep{PlottEconometrica1973}.

Any set function $S \in \mathcal S$ is \hypertarget{def mon}{\textbf{monotonic}} 
if, for each
pair $Z, Z' \in [X]^*$, $Z\subseteq Z'$ implies that $S(Z)\subseteq
S(Z')$. A choice function satisfies
\textbf{substitutes (sub)} if the rejection function associated with
it is monotonic.\footnote{This property has appeared 
  as ``Postulate 4'' in \cite{ChernoffECMA1954} and ``Property $\alpha$'' in
  \cite{SenREStud1971}. It plays 
  a central role in the matching literature under the name of
  substitutes \citep{KelsoCrawford:Econometrica1982}.}
In other words, $C \in \mathcal C$ satisfies
substitutes if, for each pair $Y, Y'\in [X]^*$,  $Y \subseteq Y'$
implies that $Y\setminus C(Y) \subseteq Y'\setminus C(Y')$.

A choice function $C \in \mathcal C$ satisfies \textbf{consistency
 (con)}, if eliminating rejected items does not affect the
choice.\footnote{Consistency has also been studied in the choice
  literature as ``Postulate 5$^*$'' in  \cite{ChernoffECMA1954} and
  ``independence of rejecting the outcast'' in
  \cite{AizermanMalishevski:IEEE1981}. Like substitutes, it plays an
  important role in the matching literature as well
  \citep{AlkanGale:JET2003,FleinerMOR2003,AygunSonmez:AER2013}. } In 
other words, for each \label{ircdef} pair $Y, Y' \in [X]^*$,
$C(Y') \subseteq Y\subseteq Y'$ implies $C(Y)=C(Y')$. 

A choice function is path independent if and only if it satisfies
substitutes and consistency \citep{AizermanMalishevski:IEEE1981}. That
is, $\mathcal C^{pi} = \mathcal C^{sub-con}$

Among path independent choice functions are those rationalized by a
preference relation and capacity. A choice function $C\in \mathcal C$
is \textbf{responsive (res)} to a linear order $\succ$ over
$X\cup\{\varnothing\}$ and quota $q$ if, for each $Y\in [X]^*$, $C(Y)$
corresponds to the $q$ highest $\succ$-ranked elements in $\{y\in Y:
Y\succ \varnothing\}$.\footnote{We denote by $\varnothing$ the option
  of not choosing an item. So, we define the order $\succ$  over
  $X\cup\{\varnothing\}$ rather than just $X$ to permit the
  possibility that some items are unacceptable.}  It is well known in
the literature that 
 $ \Cres\subset \Cpi$.

Below, we characterize the set of exclusion functions that preserve
path independence over $\Cres$. Since $\Cres \subset \Cpi$, the
restriction of the domain to $\Cres$ strengthens the necessity
results. As we will see in subsequent 
sections, however, expanding the domain  constricts the set of exclusion
functions that preserve path independence.

To ease exposition, we break up our analysis. First we separately
consider ``pure expansion'' (in the sense of exclusion functions that
are dilations) and   ``pure reuse''  (in the sense of the
exclusion function being a contraction). Then, we build on this to
handle arbitrary exclusion functions.

\subsection{Pure Expansion}
\label{sec:pure-expansion}
As explained above, we focus in this subsection on exclusion functions
that are ``expansions'' and therefore permit no reuse of chosen
items. That is, we restrict 
attention to exclusion functions that are dilations.

We start by establishing some conditions on pure expansions that are necessary for
$\mathcal L_{E}$ to preserve path independence  over $\Cres$. The first
condition is \hyperlink{def mon}{monotonicity}.
The connection between path independence of a choice function and
monotonicity of the rejection function has been well known in the literature
\citep{DanilovKoshevoyOrder2009} and is reflected in our definition of
substitutes. Even though the exclusion function
is not equivalent to the  rejection function of the lexicographic
composition, it comes as no surprise that monotonicity is a necessary
condition.

\begin{claim}\label{cl: dil monotonic}
If $E$ is a dilation and  $\mathcal L_E$ preserves
path independence   over $\Cres$, then $E$ is monotonic.
\end{claim}

A dilation $D \in \mathcal D$ is
\textbf{all-or-nothing} if, for each $Z \in [X]^*$, $D(Z)\in \{Z \cup
D(\emptyset),X\}$.\footnote{The ``all'' is in reference to the universal
  set $X$. The ``nothing'' is in reference to the value of the dilation at the
  empty set $D(\varnothing)$.} Since the set of items excluded by the empty set
 plays an important role in our analysis, we define
$K=E(\emptyset)$.

\begin{claim}\label{cl: dil all-nothing}
If $E$ is a dilation and $\mathcal L_E$ preserves path independence    over $\Cres$, then $E$ is
all-or-nothing.
\end{claim}

A dilation $D\in \mathcal D$ is \textbf{cardinal} if, for each pair
$Z, Z'\in [X]^*$, such that $Z \cup D(\emptyset)\neq X$ and
$|Z|=|Z'|$, $D(Z)=X$ implies $D(Z')=X$. 

\begin{claim}\label{cl: dil cardinal}
If $E$ is a dilation and $\mathcal L_E$ preserves path independence
over $\Cres$, then $E$ is cardinal.
\end{claim}

The next condition is equivalent to the combination of monotonicity,
all-or-nothingness, and cardinality.
A dilation $D \in \mathcal D$ is
\textbf{\boldmath threshold-linear (with threshold $t \in \mathbb N\cup\{0,\infty\}$)} if, for all $Z \in [X]^*$,

\[
  D(Z) = \left\{
    \begin{array}{ll}
Z\cup K & \text{ if }|Z|<t\\
X & \text{ otherwise}.\footnotemark
    \end{array}
  \right.
\]
\footnotetext{We use the term \emph{linearity} in the sense of taking
  the union with a constant
  set $K$. For choice functions, one could use the same terminology in
  reference to
  intersections. \cite{AizermanZavalishinPyatnitskiiARC1977a,AizermanZavalishinPyatnitskiiARC1977b} 
  study both forms of linearity but highlight their differences in
  another way. On the full domain, they show that a linear dilation
  $D$ is \textit{multiplicative} in the sense that, for all $Z,Z' \in
  [X]^*$, $D(Z \cap Z')=D(Z) \cap D(Z')$. Conversely, a linear choice
  function $C$ is \textit{additive} in the sense that, for all $Z,Z'
  \in [X]^*$, $C(Z \cup Z')=C(Z) \cup C(Z')$.}

Threshold-linearity is not only necessary but also sufficient for a dilation to preserve
path independence over $\Cres$.
\begin{proposition}\label{prop: dil cres}
if  $E$ is a dilation, then $\mathcal L_E$ preserves path independence   over
$\Cres$ if and only if $E$ is threshold-linear.
\end{proposition}

\subsection{Pure Reuse}
\label{sec:pure-reuse}
We now turn to the case of exclusions that only limit reuse and without
expanding the set. That is, we consider exclusion functions
that are contractions.

Denote by $I$
the identity exclusion function where $I(Z) = Z$ for each $Z\in [X]^*$
and by $I\setminus E$ the set function that selects $I(Z)\setminus E(Z)$ for
each $Z$.  Since we are focused on  $E$ being a contraction,
$I\setminus E$ is also a contraction. Specifically, $I\setminus E$
gives the items that are \emph{reusable} among those that have been
selected by the first choice function. We state here  conditions on
$I\setminus E$ rather than $E$ directly as this is useful when we
consider more general exclusion functions.

The first necessary condition is that if an item is reusable in some
set $Z$, then it is reusable in every superset of $Z$.
\begin{claim}
  \label{cl: reuse monotonic}
  If $E$ is a choice function and $\mathcal L_E$ preserves path independence  
  over $\Cres$, then $I\setminus E$ is monotonic.
\end{claim}

A contraction is \textbf{cardinal} if whether an item is chosen or
not depends only on the cardinality of the input set. That is, a
contraction $C$ is cardinal if for any $Z, Z'\in [X]^*$ such that
$|Z| = |Z'|$, and any
$x\in Z\cap Z'$, $x\in C(Z)$ if and only if $x\in C(Z')$.
\begin{claim}
  \label{cl: cardinal reuse}
  If $E$ is a contraction and $\mathcal L_E$ preserves path independence  
  over $\Cres$, then $I\setminus E$ is cardinal.
\end{claim}

The next condition is equivalent to the combination of  monotonicity and cardinality of
$I\setminus E$. A contraction
$C\in \mathcal C$ is \textbf{cardinal-linear} if there is a
 sequence $\{T^n\}_{n=0}^\infty$ such that
$\emptyset = T^0 \subseteq T^1\subseteq T^2\dots$,  and for each $Z\in
[X]^*$, $C(Z) =  Z\cap T^{|Z|}$.

\begin{proposition}\label{prop: contr cres}
If $E$ is a contraction, then $\mathcal L_E$ preserves path
independence over $\Cres$ if and only if $I\setminus E$ is
cardinal-linear. 
\end{proposition}

\subsection{Putting the Pieces Together}
\label{sec:putt-piec-togeth}
So far, we have considered two extremes of how an exclusion might
behave: pure expansion and pure reuse.

An arbitrary exclusion function $E\in \mathcal S$ can be
decomposed into two such parts. That is, we can express it as the 
difference between a dilation $G_E \in \mathcal{D}$ and a contraction
 $R_E \in \mathcal{C}$ where, for each $Z 
\in [X]^*$
\[
E(Z)=G_E(Z)\setminus R_E(Z)
\]
with $G_E(Z)=E(Z) \cup Z$ and $R_E(Z)=Z \setminus E(Z)$. 

These two components of the decomposition are
economically meaningful. Intuitively, $R_E$ captures the scope of
\textbf{reuse} by the second choice function $C_2$. If $R_E(Z) \neq
\emptyset$, then some items  chosen by the first choice
function $C_1$ can be rechosen by $C_2$. In turn, $G_E$ captures the
scope of \textbf{gross exclusions} that cannot contribute to
\textit{incremental} choice by $C_2$. For each $z\in G_E(Z)$,  either
$z \in E(Z)$ so that $z$ 
is not choosable by $C_2$; or $z \in R_E(Z)$ so that $C_2$ can choose
$z$ but cannot add it to
the aggregate choice (as it is already chosen).

If $E$ is a contraction, then since the domain of $E$ is all
finite subsets of $X$, it is never the case that $G_E(Z) = X$ for any
$Z$. However, if $E$ is not a contraction, whenever $G_E(Z) = X$, $R_E$
is irrelevant. In that
case, the set of items from which $C_2$ must choose is a subset of
$Z$, the items already chosen by $C_1$. So, whatever  scope of reuse
is permitted, there is no way for $C_2$ to  affect the aggregate
choice. Denote the domain of sets where  
$R_E$ is (potentially) \textbf{relevant} by $Dom(R_E)=\{Z \in [X]^*:
G_E(Z)\neq X\}$.\footnote{If $G_E$ is monotonic and $Z\in Dom(R_E)$,
  then for all $Z'\subset Z$,  $Z'\in Dom(R_E)$. Moreover, if $G_E$ is
  all-or-nothing, then $G_E(Z) = Z\cup K \subset X$ for all $Z\in
  Dom(R_E)$.}

The above observation about relevance leads us to refine our
necessary conditions from \Cref{sec:pure-reuse}.
Given an exclusion function $E\in \mathcal S$,  $R_E$
is \textbf{monotonic on \boldmath $Dom(R_E)$} if for any $Z,Z'\in
Dom(R_E)$ such that $Z\subseteq Z'$, $R_E(Z) \subseteq
R_E(Z')$. Similarly, $R_E$ is \textbf{cardinal-linear on  \boldmath
  $Dom(R_E)$} if there is a 
 sequence $\{T^n\}_{n=0}^\infty$ such that
$\emptyset = T^0 \subseteq T^1\subseteq T^2\dots$,  and for each $Z\in
Dom(R_E)$, $R_E(Z) =  Z\cap T^{|Z|}$. The only difference from our
earlier definitions is that we only require the conclusion to hold
over $Dom(R_E)$ as opposed to all of $[X]^*$.

The  necessary conditions in
\Cref{sec:pure-expansion} and the above adaptations of those in \Cref{sec:pure-reuse} are necessary for
general exclusion functions as well, except that they apply to the
corresponding gross exclusion and reuse components.

\begin{claim}
  If $\mathcal L_E$ preserves path independence  over $\Cres$, then
  \begin{enumerate}
  \item  $G_E$ is threshold-linear, and 
  \item  $R_E$ is cardinal-linear on $Dom(R_E)$.
  \end{enumerate}
\label{cl: general nec cres}  
\end{claim}

The conditions listed in \cref{cl: general nec cres} leave out the
interaction between the gross exclusion and reuse. The next condition
says that no items in $K$ can ever be reused.
Given an exclusion function $E\in\mathcal S$, $R_E$ is
\textbf{\boldmath $K$-disjoint on $Dom(R_E)$} if for every $Z\in
Dom(R_E)$, $R_E(Z) \cap K = \emptyset$.

\begin{claim}
  \label{cl: k disjoint}
  If $\mathcal L_E$ preserves path independence over $\Cres$, then $R_E$ is
  $K$-disjoint on $Dom(R_E)$. 
\end{claim}

We now define a class of exclusion functions that satisfy all of the
above mentioned necessary conditions to preserve path independence.
An exclusion function $E\in \mathcal S$ is \textbf{threshold-linear
  with cardinal reuse} if there are $t\in \mathbb N\cup\{0,\infty\}$,
$K\subseteq X$, and $\{T^n\}_{n=0}^t$ where $\emptyset = T^0
\subseteq T^1\subseteq \dots \subseteq T^t \subseteq X\setminus K$,
such that for each $Z\in [X]^*$, if $|Z|< t$
\[
  E(Z) = (Z\setminus T^{|Z|})\cup K,
\] and otherwise $G_E(Z) = X$.
It follows that $Dom(R_E) = \{Z\in [X]^*: Z\cup K \subset X\text{ and
}|Z|<t\}$ and for each $Z\in Dom(R_E)$, $G_E(Z) = Z\cup K$ and $R_E(Z)
= Z\cap T^{|Z|}$.

\begin{theorem}
  \label{thm: cres}
  $\mathcal L_E$ preserves path independence over $\Cres$ if and only
  if $E$ is threshold-linear with cardinal reuse.
\end{theorem}

Note that \Cref{thm: cres} implies both \Cref{prop: dil cres} and
\Cref{prop: contr cres}. For the former, we simply restrict attention
to exclusion functions that are threshold-linear with cardinal reuse
where $T^l = \emptyset$ for each $l$. For the latter, we restrict
attention to the identity dilation, $G_E(Z) = Z$ for every $Z\in
[X]^*$.

By \Cref{thm: cres}, the preservation of path independence over
$\Cpi$ imposes a considerable amount of structure on the exclusion
function. How much of this is due to the substitutes component of path
independence as opposed to consistency? All of it: the preservation of
consistency, even on the larger domain of $\Ccon$ places no
constraints on the exclusion function.
\begin{remark}
  For any $E\in \mathcal S$, $\mathcal L_E$ preserves consistency over $\Ccon$.
\end{remark}
\section{Broader Domains and Additional Properties}
\label{sec:addit-prop-broad}
There are two pertinent ways in which we may modify the question
answered by \Cref{thm: cres}. The first is to extend the domain beyond
$\Cres$. By considering the preservation of path independence over
$\Cpi$, we can accommodate nested composition of more than two choice
functions, which we return to in \Cref{sec:more-than-two}.  In
the second direction, we consider the preservation of  more than just path
independence. 

\subsection{Expanding the Domain}
\label{sec:consistency-csubirc}

While threshold-linearity with cardinal reuse ensures that an
exclusion function preserves path independence over $\Cres$, such a
guarantee does not hold for input choice functions that are path
independent but not responsive.\footnote{Note that if we take inputs
  that are not themselves  path independent, there is no hope that the
  composite choice function would be path independent. For this reason
  we stop at expanding the domain to $\Cpi$.}

Since $\Cres\subset \Cpi$, \Cref{thm: cres} implies that it is
necessary for any $E\in \mathcal S$ such that $\mathcal L_E$ preserves
path independence over $\Cpi$ to be threshold-linear with cardinal
reuse. However, this is not sufficient as only some of these exclusion
functions preserve path independence over the larger
domain. The only threshold-linear exclusion functions with cardinal
reuse that do so  are ones that are almost constant (the threshold is
either 0 or 1) or are linear (the threshold is $\infty$). Moreover,
reuse is limited in that expanding the choice beyond two items  does
not induce further reusable items.

\begin{proposition}
  \label{prop: pi}
  $\mathcal L_E$ preserves path independence over $\Cpi$ if and only
  if $E$ is threshold-linear with cardinal reuse with respect to
  $K\subset X$, $t\in \{0, 1, \infty\}$, and $\{T^n\}_{n=0}^t$ such
  that if $t > 0$ then  $\emptyset = T^0 \subseteq T^1\subseteq T^2 =
    T^3 \dots = T^t\subseteq X\setminus K$. 
\end{proposition}

If we further expand the domain to $\Csub\supset \Cpi$, only the
possibility of $t=\infty$  with $T^1 = T^2 \dots =
T^\infty$ remains if we wish to preserve substitutes.

\subsection{Preserving an Additional Property}
\label{sec:size-monot-csubsm}
We now see how the result changes if we are more demanding, not about
the domain, but about what properties are preserved. We consider an additional property
that says the choice from a
set contains at least as many elements as a choice from any subset.
A choice function $C \in \mathcal C$ is \textbf{size monotonic (sm)}
if, for each pair $Y, Y'\in [X]^*$, $Y \subseteq Y'$ implies that
$|C(Y)| \leq |C(Y')|$.  This property arises often in the matching
literature along with substitutes and the following is well known.

\begin{remark}
$\Cres \subseteq \Csubsm \subset \Cpi$.
\end{remark}

We now consider the preservation of not only path independence, but
the conjunction of substitutes and size monotonicity over
$\Cres$.\footnote{The sufficiency part of \Cref{thm: cres} actually
  extends to $\Csubsm$. That is, if $E\in \mathcal S$ is
  threshold-linear with cardinal reuse, then $\mathcal L_E$ preserves
  path independence over $\Csubsm$. However, as stated in \Cref{prop:
    sm} threshold linearity with cardinal reuse is not sufficient to
  preserve size monotonicity. }
The additional requirement shrinks the set of exclusion functions,
much as expanding 
the domain did in \Cref{sec:consistency-csubirc}. 
The added exigency of preserving size monotonicity not only  restricts
the threshold, but  also severely limits reuse.

\begin{proposition}
  \label{prop: sm}
  $\mathcal L_E$ preserves substitutes and size monotonicity over
  $\Cres$ if and only if
   $E$ is threshold-linear with cardinal reuse with respect to
   $K\subset X$, $t\in \mathbb N\cup\{0,\infty\}$, and
   $\{T^n\}_{n=0}^t$ such that either 
  \begin{enumerate}
  \item $|X\setminus K| \leq 1$ or 
  \item $t=\infty$ and for each $n < t$,  $T^n = \emptyset$. 
  \end{enumerate}
\end{proposition}

\section{Nested Composition}
\label{sec:nested}
Given an exclusion function $E\in \mathcal S$, $\mathcal L_E$ is a
binary operator on choice functions. For special cases such as the
identity exclusion function, it is associative. In fact, for the empty
exclusion (``full reuse'' as in the representation of
\cite{AizermanMalishevski:IEEE1981}) it is 
 commutative as well. However, $\mathcal L_E$ is not generally
associativity or commutative. In this section, we consider a few ways
in which compositions may be nested and  pay special attention to
the aggregation of
single valued choice functions.

\subsection{Composing More Than Two Choice Functions}
\label{sec:more-than-two}
As mentioned above,  lexicographic composition is not
generally associative or commutative. Since the purpose of
composition is to build up ``larger'' choice functions from  various
``parts,'' there are many ways we can go about this. To illustrate
what is at stake, we start with just three choice functions $C_1,
C_2,$ and $C_3$ in that order. In this case, there are two possibilities, which can
be thought of as  ``left composition'' and  ``right composition,'' 
respectively:
\[
  \mathcal L_{E_1}  (C_1, \mathcal L_{E_2}(C_2, C_3)) \text{ and }
  \mathcal L_{\hat E_2}  (\mathcal L_{\hat E_1}(C_1, C_2), C_3) \]
Where $(E_1, E_2)$ and $(\hat E_1, \hat E_2)$ are two pairs of
exclusion functions.

Neither direction of composition is more general than the other. To
see this, consider what happens  if $E_1 = \hat E_1$ and $E_2 = \hat E_2$, and
we start from some set $Y$ to choose from. In the first step, $C_1$
selects $Z_1 = C_1(Y)$ no matter which of the two directions we
compose in. The set that $C_2$ chooses from is then $Y\setminus
E_1(Z_1)$ in both cases. Thus, the inputs to the first two choice
functions are exactly the same, so $C_2$ selects $Z_2 = C_2(Y\setminus
E_1(Z_1))$. The difference, however, is when we
get to the third choice function.
The exclusion  from the set that  $C_3$ chooses from, under left and
right compositions, are
\[ E_2\left[
    Z_1 \cup Z_2
  \right] \text{ and }
  E_1(Z_1) \cup E_2(Z_2)
\]  respectively. These are not necessarily the same. As we show below, neither format
is more general than the other in terms of what
policies can be achieved.  Suppose $C_1, \dots C_m$ are a sequence of
choice functions where $m>2$. Consider two procedures aggregating
these $m$ choice functions.

\begin{procedure}\label{proc:aggregate-quota}
{\em  Fix an integer $N$. Starting with $C_1$ and continuing until $C_m$,
  each successive 
  choice function can choose freely from the remainder of the  input set if
  prior choice functions selected fewer than a total of $N$
  alternatives, but it cannot choose anything otherwise.}
\end{procedure}

\begin{procedure}\label{proc:individual-quota}
{\em  Fix an integer $N$. Starting with $C_1$ and continuing until $C_m$,  each successive
  choice function can choose freely from the remainder of the  input
  set if 
  each prior choice function \emph{individually} selected fewer than $N$
  alternatives, but it cannot choose anything otherwise.}
\end{procedure}
\begin{claim}\label{cl:left-right-comp} 
  \begin{enumerate}
  \item   \Cref{proc:aggregate-quota} can be implemented via right composition
  but not left composition. 
\item   \Cref{proc:individual-quota} can be implemented via left composition
  but not right composition. 
  \end{enumerate}
\end{claim}

These two procedures capture the crucial difference between right and
left composition. Left composition cannot condition exclusion for the
$k^{\text{th}}$ choice function on anything other than what the
$k-1^{\text{th}}$ choice was. In this sense, it has ``no memory'' of
prior choices. While prior choices still  affect
  what is available to the $k^{\text{th}}$ via the input set to not
  only the  $k^{\text{th}}$ but also the  $k-1^{\text{th}}$ choice
  function, they do not affect it through exclusion.  On the other
  hand,  right composition necessarily conditions exclusion for
  the $k^{\text{th}}$ choice function on the \emph{union} of all prior
  choices. In this sense, there is an ``aggregate memory'' of prior choices
  (though the individual choices are not discernable).  Both  \emph{no
  memory} and \emph{aggregate memory} are restrictive. However, as we have
shown  above, they are restrictive in different ways.

With more than three choice functions, one can combine them by mixing
right and left composition. Specifically there are
$\frac{1}{m}{2(m-1)\choose m-1}$ (the $m-1^{\text{th}}$ Catalan
number) ways to combine  $m$ choice functions. Below are a few
examples.
\begin{example}{\em 
    \textbf{Soft quotas.} Suppose $X$ is the universe of candidates a firm
    with $n$ divisions can choose to  hire from. Let $C_i$ be the
    $i^{\text{th}}$ division's choice function. Moreover, suppose that
    each hire can only work at one division. The firm can impose a
    soft quota of $k$ on the number of hires as follows. For each
    $Z\in [X]^*$, let $E(Z)$ be  $Z$ if  $|Z| \leq k$ and $X$
    otherwise. The aggregated choice function is then
    \[
      \mathcal L_E(\mathcal L_E(\dots \mathcal L_E(C_1, C_2),\dots), C_n).
    \]
    Then, for any set of applicants $Y$, the divisions get to choose
    whom to hire, from the first to the $n^{\text{th}}$. However, the
    process stops as soon as at least $k$ applicants have been chosen.
    
    If the component choice functions $C_i$ are known to be single-valued, $k$
    serves as a hard constraint.
  }
\end{example}

\begin{example}\label{ex:reserve}
  {\em
    \textbf{Nested reserves.}
    Let  $\{C_j\}_{j=1}^n$ be a sequence of $n$
    choice functions and $\{X_j\}_{j=1}^n$ be a monotonic sequence of
    subsets of 
    $X$ such that $X_1\supseteq X_2 \supseteq \dots \supseteq X_n$. Each
    item in $X_j$ can only be chosen by $C_i$ such that $i\leq
    j$. Thinking of the choice functions as representing resources,
    the $j^{\text{th}}$ resource is reserved for $X_j$. Thus, elements
    of $X_n$ are most favored while elements of $X_1\setminus X_2$ are
    least favored.
  }
\end{example}
\begin{example}{\em
    \textbf{Inter-district school choice.}
    In the school choice model, $X$ is the set of all
    students. Reserves like in \Cref{ex:reserve} are a common way to
    implement policies like affirmative action in schools' choice
    functions. The
    inter-district version of this model
    \citep{HafalirKojimaYenmez2018} is nested in the sense  the
    schools' choice functions are aggregated into a district level
    choice functions and  students are then matched to districts.    
    }
\end{example}

\subsection{Single-valued Inputs}
\label{sec:single-valued-inputs}
When it comes to constructing choice functions from simpler
components, \textbf{single-valued, responsive choice functions (sv-res)} are a salient
domain.\footnote{By ``single-valued'' we mean that $q = 1$,  so that
  such a choice function may select $\emptyset$.} They are commonly
studied in the matching literature and  
often composed using the identity
exclusion function \citep{KominersSonmezTE2016}. At the opposite
extreme, lexicographic composition of such choice functions with the
empty exclusion function spans $\Cpi$
\citep{AizermanMalishevski:IEEE1981}. In this section, we consider the
preservation of path independence when lexicographically combining a
single-valued, responsive choice function  with a path independent
choice function.

We first consider lexicographic composition with the first choice
function being in $\Csvres$ and the second in $\Cpi$. The only
relevant parts of an exclusion function's domain in this case are the
set of singletons and the empty set.

\begin{proposition}
    \label{prop:sv-sub}
    $\mathcal L_E$ preserves path independence over $\Csvres\times 
    \Cpi$ if and only if, on the set of singletons and the empty set
    $E$ coincides with an exclusion function that is threshold-linear
    with cardinal reuse.
  \end{proposition}

Next, we consider preserving path independence over $\Cpi\times
\Csvres$. The proofs of necessity building up to (and
including) that of \Cref{prop: pi}
do not use the full domain of $\Cres$ for the second choice
function. They only appeal to single-valued choice
functions. Sufficiency of the same conditions, of course, follows from
sufficiency for the broader domain of $\Cpi$.

\begin{proposition}
  \label{prop:sub-sv}
    $\mathcal L_E$ preserves path independence over $\Cpi\times
    \Csvres$ if and only
  if $E$ is threshold-linear with cardinal reuse with respect to
  $K\subset X$, $t\in \{0, 1, \infty\}$, and $\{T^n\}_{n=0}^t$ such
  that if $t > 0$ then  $\emptyset = T^0 \subseteq T^1\subseteq T^2 =
  T^3 \dots = T^t\subseteq X\setminus K$.
\end{proposition}

In \Cref{apx:sm-sv},  we consider the preservation of size
monotonicity as well. Since applications typically involve nested
composition, we restrict the domain to  $ \Csvres \times \Csubsm$ for
the analog of \Cref{prop:sv-sub} and to   $\Csubsm\times \Csvres$ for
the analog of \Cref{prop:sub-sv}.

\section{Equivalence Relations  and Matching With Contracts }
\label{sec:equiv}
In the canonical setting of
many-to-one 
matching with contracts, the market participants are hospitals and
doctors. While a hospital may choose multiple contracts, it cannot
choose more than one contract per doctor. In this context, when the
items are contracts, there is an equivalence relation over items: Are
$x$ and $y$ contracts with the same doctor?
We can then  consider choice functions
that select no more than one element of each equivalence class. 

Denote by $\sim$ the equivalence relation in question. 
We assume that the number of equivalence classes of $\sim$ is  countable. Given 
$x\in X$, let $I_x$ be the equivalence class of $\sim$ that $x$
belongs to. Given a set $Z\subseteq X$, let $I_Z = \cup_{x\in Z}I_x$.

The choice function $C$ is \textbf{many-to-one} (mto1) if, for each $Y
\in [X]^*$, $x,z \in C(Y)$ implies $x\nsim z$. In
this setting, the natural  minimal exclusion function forces the
composition of two choice functions to respect many-to-oneness. We
define it as follows. For each $Z \in [X]^*$,
\[
  \underline E(Z) = I_Z.
\]

Using this definition, our notion of lexicographic composition
$\mathcal L_{\underline E}$ generalizes the \textit{slot-specific
  priorities} model of \cite{KominersSonmezTE2016}. They showed that
$\mathcal L_E$ need not preserve substitutes over $\Cres$ but must
be  consistent and satisfy a weaker ``bilateral''
notion of substitutes first proposed by
\cite{HatfieldKojima:JET2010}. Subsequently, \cite{HatKom:2014HSubs}
showed that any 
choice function with slot-specific priorities can be ``completed'' so
that it satisfies substitutes and size monotonicity. 

Let $[X]^*_F=\{ Y \in [X]^* : |\{I_x\}_x\in Y|=|Y|\}$ be the
\textbf{feasible sets} of items that contain no more than one item
from each equivalence class. Then, $\overline C \in \mathcal C$
\textbf{completes} a 
many-to-one choice function $C \in \mathcal C^{\text{mto1}}$ if, for all $Y \in
[X]^*$, $\overline C(Y) \in [X]^*_F$ implies $\overline C(Y) = C(Y)$. 

Analogous to earlier notation, for a given property $\pi$, denote by
$\overline {\mathcal C}^\pi$ the set of many-to-one choice functions
with completions that satisfy $\pi$.

To preserve many-to-oneness of the input choice functions, it is
necessary (and sufficient) that, for each set, $E$ extends $\underline
E$. Formally, an exclusion function $E\in \Ens$ is
\textbf{equivalence-excluding} if, for all $Z \in [X]^*_F$, $Z\cup E(Z) \supseteq
\underline E(Z)$. 

\begin{remark}\label{rmk:many-one} 
$\mathcal L_E$ preserves many-to-oneness over $\mathcal C^{\text{mto1}}$ if and
only if $E$ is equivalence-excluding. 
\end{remark}

Inspection of the proof makes it clear that this same condition is
necessary (and sufficient) to preserve many-to-oneness for input
choice functions in $\Chres$.

For the exclusion function $\underline E$, which often
appears  in the literature, while  $G_{\underline E}$ is  monotonic, it is
\textit{not} all-or-nothing. Indeed for any equivalence-excluding
exclusion function $E$, if $G_E$  is
all-or-nothing then for every nonempty $Z\in [X]^*$ $G_E(Z) = X$, implying that
$Dom(R_E) = \{\emptyset\}$. This means that, to preserve path independence \emph{and}
many-to-oneness (over $\Chres$), the exclusion
function $E$ must entirely shut down the second input $C_2$. 

Nonetheless, simple adaptations of our earlier conditions
are sufficient to preserve the possibility of a path independent
\emph{completion} over $\Chpi$
and  size monotonicity over 
$\Chsubsm$.

As long as the first input choice function is many-to-one,
the effective domain of $E$ is the range of a \mto choice
function. So, there is no need to impose any restrictions on $E$
beyond $[X]^*_F$.

With this in mind, we modify our definition of threshold-linearity
with cardinal reuse. An exclusion function $E\in \mathcal S$ is
\textbf{many-to-one threshold-linear with cardinal reuse} if there are
$t\in \mathbb N\cup \{0,\infty\}$, $K\subseteq X$, and
$\{T^n\}_{n=0}^t$ where $\emptyset  = T^0 \subseteq T^1 \subseteq
\dots \subseteq T^t \subseteq X\setminus K$, such that \underline{for each $Z\in
[X]^*_F$}, if $Z\cup K \subseteq X$ and $|Z|< t$, $E(Z) = (Z\setminus
T^{|Z|})\cup K$ and otherwise $G_E(Z) = X$. 
The only change from the definition of threshold-linearity with
cardinal reuse in \Cref{sec:path independence} is the underlined part.

Analogs to the sufficiency parts of \Cref{thm: cres,prop: pi,prop: sm}
are consequences of  the following result.
\begin{lemma}
  \label{lem:mto1}
    Suppose that  $E$ is   equivalence-excluding and many-to-one
    threshold-linear with cardinal
  reuse for parameters $t,K,$ and $\{T^s\}$. Let $\overline E$ be
  threshold-linear with cardinal reuse with the same parameters. 

  If  $\overline{C_i}$ completes $C_i$ for $i=1,2$ and, in addition,
  $\overline{C_2}$ satisfies \IRC, then $\mathcal L_{\overline
    E}(\overline{C_1},\overline{C_2})$ completes $\mathcal
  L_E(C_1,C_2)$. 
\end{lemma}
It is natural to think of $E$ as ``removing'' items from what is
available to $C_2$ as a function of what $C_1$
chooses. As stated in \Cref{rmk:many-one}, preserving many-to-oneness
\emph{requires} that $E$ remove \emph{all} items related by $\sim$ to
any item chosen by $C_1$. The idea of the exclusion
functions $\overline E$ defined in \Cref{lem:mto1} is to ``put
back'' all but the items  that were actually chosen by $C_1$. 

The necessary conditions, however, are not exactly analogous to those
in  \Cref{thm: cres,prop: pi,prop: sm}.
In particular,  the all-or-nothingness of gross exclusion  is \textit{not}
necessary to preserve path independent  completability over $\Chpi$ (even
though the standard version of this condition \textit{is} necessary to
preserve path independence over $\Cpi$). 

\begin{example}\label{ex:all or nothing not necessary}
  {\em
    Let $X$ be such that $X \supseteq \{a,b,c\}$ where $a\sim c \nsim
    b$ and $I_a = \{a,c\}$. That is, items $a$ and $c$ are related by
    $\sim$ and complete their equivalence class, and there is a third
    item  $b$ that is unrelated to $a$ and $c$.  Let $E$ be such that
    for each $Z\in [X]^*_F$, 
    \[
      E(Z) = \left\{
        \begin{array}{ll}
          X\setminus \{a\}&\text{if }I_a \cap I_Z = \emptyset,\\
          I_a & \text{if }I_a = I_Z ,\\
          X & \text{if } I_a \subsetneq I_Z.
        \end{array}
        \right.
    \]
Note that $E$ is equivalence-excluding and $G_E$ is monotonic over
$[X]_F^*$. However, it 
violates all-or-nothingness  since, for instance, $\{b\}\cup E(\{b\}) =
X\setminus \{a\}$ while $\{a,b\}\cup E(\{a,b\}) = X$. 
    Given a pair $C_1, C_2\in \Chpi$, let $C = \mathcal L_E(C_1,C_2)$. Let $\overline C\in \Csub$ be such that for each $Y\in [X]^*$,
    \[
      \overline C(Y) = \left\{
        \begin{array}{ll}
          C(Y)& \text{if } Y\neq \{a,c\},\\
          \{a,c\} & \text{if }a\in C(X) \text{ and } C(\{a,c\}) = \{c\},\text{ and}\\
          C(\{a,c\}) & \text{otherwise. }\\
        \end{array}
        \right.
      \]
      $\overline C$ is a path independent completion of $C$ even though
      $G_E$ violates  all-or-nothingness on $[X]_F^*$.   }
  \end{example}
  
Our  reasoning from the proof of  \Cref{thm: cres} does not hold here
because the sets picked by the first choice function are not
exhaustive and are limited to $[X]^*_F$. The  
way \Cref{ex:all or nothing not necessary} deviates from it is, in a
sense, the limit. In \Cref{apx:weak-AorN}, we  weaken
all-or-nothingness to allow only the violations such as in
\Cref{ex:all or nothing not necessary} and show that such a condition
is necessary.

\section{Discussion}
\label{sec:discussion}

We have considered choice functions of the type $[X]^*\to [X]$. Our
formulation of feasibility as exclusion is appropriate for 
applications where the designer of the aggregation rule does not
control the individual choice functions. In a sense, our results
delineate the limits of what such a designer can achieve.

What if, instead, the designer can exert more control on the
individual choice functions?  Perhaps a choice function takes as input not only a
set that it may choose from, but also parameter from the set $\mathcal
A$. A choice function would then be of the type $[X]^*\times \mathcal
A \to [X]$. An exclusion function, in this case, would also be of this
type. Since the designer of the exclusion functions would have more
information to condition exclusion on, as well as an extra parameter to
the individual choice functions, they would have greater influence on
the  aggregate choice function.

As  an extreme example, the parameters could include all of the
information generated by the composition process in the form of a
sequence of sets: the initial
set, the set chosen by the first choice function, that chosen by the
second choice function, and so on. This is a strictly more general
formulation than ours. In our formulation, if $C_2$ does not see an
item $x$ in its input, it cannot tell whether this is because $x$ was
never a possibility or because it has been excluded based on earlier
choices.
In the more general model, the $k^{\text{th}}$  choice function $C_k$ would
have as an extra input a
sequence  of sets of items $\{Z_n\}_{n=0}^{k-1}$ where $Z_0$ is the
initial set and $Z_1$ through $Z_{k-1}$ are the choices of the first
through $k-1^{\text{th}}$ choice functions. The exclusion $E_{k-1}$ applied
prior to the $k^{\text{th}}$ choice, would take the same sequence as an
input and return a menu of sets that $C_k$ can choose from. The right
composition of a sequence of $n$ choice functions would 
then  produce from the input $Y$ the choice $\cup_{i=1}^n Z_i$ where
$Z_0 = Y$, $Z_1 = C_1(Y, \{Z_0\})$, and for each $i = 2,\dots, n$,
\[
  Z_i = C_i(Y\setminus E_i(\{Z_l\}_{l=0}^{i-1}), \{Z_l\}_{l=0}^{i-1})
\]
For such a model, 
the question would be \emph{``what conditions on $E$ and the way
  the $C_i$s depend on the parameter ensure that the resulting
  choice function inherits desirable properties of the individual 
  choice functions?'' }
A comprehensive analysis of this more general question  is beyond the
scope of the current paper, which is a first step towards
understanding path independence of lexicographic compositions.

A tractable next step could be to focus on specific features of prior
choices. For instance, contrary to our definition
of $\Cres$, where
capacities are inherent to the choice functions, they might be a
parameter. Exclusions could depend on
the cardinalities of prior choices.
 \cite{WestkampET2013} demonstrates
  a  way in which the cardinalities along these sequences can
  factor into the choices so that the responsive choice functions
  yield a path independent choice function in the
  end.\footnote{\cite{AygunTurdhan2018} just extend this to
    equivalence-based setting like in \Cref{sec:equiv}.}

\appendix

\section{Proofs}
\label{apx:proofs}

\subsection{Proofs From  \Cref{sec:path independence}}
\Cref{cl: dil monotonic,cl: dil all-nothing,cl: dil cardinal,cl: reuse monotonic,cl: cardinal reuse} are implied by  \Cref{cl: general nec cres}.
\Cref{prop: dil cres,prop: contr cres} are special cases of
\Cref{thm: cres}. So, we omit the proofs of the results from
\Cref{sec:pure-expansion,sec:pure-reuse} and proceed to the proofs
of results  in \Cref{sec:putt-piec-togeth}.

\begin{proof}[{\bf\em Proof of \Cref{cl: general nec cres}}] To
  conclude that $G_E$ is threshold-linear, we prove that it satisfies
  each of the properties defined in \Cref{sec:pure-expansion}.

\noindent \textbf{Monotonicity of \boldmath $G_E$:}
  Suppose that $G_E$ is not monotonic. Then, there are $Z,Z'
\subset X$ such that $Z\subseteq Z'$ and $G_E(Z)\not\subseteq
G_E(Z')$. Let
$a\in G_E(Z)\setminus G_E(Z')$. Since $Z\subseteq Z'$,
$a\notin Z$ and hence $a\in E(Z)\setminus E(Z')$.

Let $C_1\in \Cres$ 
be induced by  $\succ_1$ such that $\{x\in X:x\succ_1 \varnothing\} = Z'$
and  $q_1=
|Z'|$.
Let $C_2\in \Cres$ be
induced by
$\succ_2$ such that
$\{x\in X:x\succ_2 \varnothing\} = \{a\}$ and $q_2= 1$. 

Let $C = \mathcal L_E(C_1,C_2)$. Then, $a \in C(Z'\cup\{a\})
= Z' \cup \{a\}$ but $a \notin C(Z\cup\{a\}) = Z$, which violates
path independence. So, $G_E$ is monotonic.

\noindent \textbf{All-or-nothingness of \boldmath $G_E$:} Suppose that $G_E$ is monotonic but not all-or-nothing. Let
$\underline Z \in [X]^*$ be such that $X\supset G_E(\underline Z)
\supset \underline Z \cup K$ and for each $Z\subset
\underline Z$, $G_E(Z) = Z\cup K$.
Since $K = E(\emptyset) = G_E(\emptyset) = \emptyset\cup K$, such
$\underline Z$ exists.
By definition of $\underline
Z$, there is some $a \in  X\setminus G_E(\underline Z)$,
$b \in \underline Z$, and $c \in E(\underline Z) \setminus (\underline
Z \cup K)$.

Let $C_1\in \Cres$
be induced by $\succ_1$ such that
$\{x\in X:x\succ_1 \varnothing\} =  \underline Z$
and  $q_1 = |\underline
Z|$.
Let $C_2\in \Cres$ be induced by 
$\succ_2$
such
that  $\{x\in X:x\succ_1 \varnothing\} =  \{a,c\}$ and  $c\succ_2 a
\succ_2 \varnothing$ and $q_2 = 1$.

Let $C = \mathcal L_E(C_1,C_2)$. Then, $a \in C(\underline Z \cup
\{a,c\}) = \underline Z \cup \{a\}$ and $a \notin C((\underline Z \cup
\{a,c\})\setminus \{b\}) = (\underline Z \cup \{c\}) \setminus \{b\}$,
which violates path independence. So, $G_E$ is all-or-nothing.

\noindent \textbf{Cardinality of \boldmath$G_E$:}
Let $Z, Z'\in [X]^*$ be such that $|Z| = |Z'|$, $Z\cup K\neq X$, and
$G_E(Z) = X$. We consider two cases to show that this implies that
$G_E(Z') = X$.
\begin{description}
\item[Case 1 {\boldmath($Z\cup Z'\cup K \subset X$)}:]
Towards a contradiction, suppose that $G_E(Z') \neq X$. Since $G_E$ is
all-or-nothing and monotonic, $G_E(Z'') = Z''\cup K$ for all $Z''
\subseteq Z'$. 

Let $a \in X\setminus (Z\cup Z' \cup K)$.

Let $C_1\in \Cres$ be induced by
$\succ_1$ such that $\{x\in X:x\succ_1
\varnothing\} = Z\cup Z'$ and, for each $z' \in Z'$ and $z\in
Z\setminus Z'$, $z'\succ_1 z$
and $q_1 = |Z|$.
Let $C_2\in \Cres$ be induced by $\succ_2$ such that $\{x\in
X:x\succ_2 \varnothing\} = \{a\}$  and $q_2 = 1$. Let
$C= \mathcal L_E(C_1,C_2)$. By definition, $a\in C(Z \cup Z' \cup
\{a\}) = Z' \cup \{a\}$ and $a\notin C(Z \cup \{a\}) = Z$, which
violates substitutes. So, $G_E(Z') = X$. 

\item[Case 2 (\boldmath $Z\cup Z'\cup K = X$):]  Let $a\in X\setminus
  (Z \cup K)$ and $b \in X \setminus (Z \cup \{a\})$. Since $Z \cup Z'
  \cup K =X$, $a \in Z'$. Let $Z''= (Z'  \cup \{b\}) \setminus
  \{a\}$. Since 
$Z \cup Z'' \cup K \subset X$, $G_E(Z)=X$ implies $G_E(Z'')=X$ by the
argument in Case 1. (This argument holds even if $b \in Z'$.) If $b
\notin Z'$, then $G_E(Z'')=X$ implies $G_E(Z')=X$ by the argument in Case
1. Otherwise, $b \in Z'$. Then, $G_E(Z'')=X$ implies $G_E(Z')=X$ by
monotonicity of $G_E$.
\end{description}
Since $G_E$ is monotonic, all-or-nothing, and cardinal, it is threshold-linear.
Before showing that $R_E$ is cardinal-linear on $Dom(R_E)$, we show
that it is monotonic on $Dom(R_E)$.

\noindent\textbf{Monotonicity of \boldmath $R_E$ on $Dom(R_E)$:}
Suppose $R_E$ is not monotonic on $Dom(R_E)$. Then, there are  $Z, 
Z'\in Dom(R_E)$ 
such that $Z\subset Z'$ and $R_E(Z) \not\subseteq R_E
(Z')$. So, there is some $a\in Z$ such that $a\in R_E(Z)\setminus
R_E(Z')$. Since $Z'\in Dom(R_E)$, there is some $b\in X\setminus
G_E(Z')$. (Since $b \notin E(Z')$ but $a\in E(Z')$,  $b\neq
a$.) 

Let $C_1\in \Cres$
be induced by $\succ_1$ such that
$\{x\in X:x\succ_1 \varnothing\} =  Z'$ and $q_1 = |Z'|$. Let $C_2\in
\Cres$ be induced by 
$\succ_2$ such
that  $\{x\in X:x\succ_2 \varnothing\} =  \{a,b\}$ and  $a\succ_2 b
\succ_2 \varnothing$ and $q_2 = 1$. 

Let $C= \mathcal L_E(C_1,C_2)$. Then, $b \in C(Z' \cup \{b\}) =
Z'\cup\{b\}$ but $b \notin C(Z \cup \{b\}) =Z$, which
violates path independence. So, $R_E$ is monotonic on $Dom(R_E)$.

\noindent\textbf{Cardinal-linearity of \boldmath$R_E$ on $Dom(R_E)$:} 
We first show that there is no   pair
$Z,Z'\in  Dom(R_E)$ such that $|Z| = |Z'|$ for which there is   $a\in Z\cap Z'$ such
that $a\in R_E(Z) \setminus R_E(Z')$. If such a pair does exist, there
are two cases to consider.
\begin{description}
\item[Case 1 (\boldmath$K\cup Z\cup Z' \neq X$):] Let $b\in X\setminus
  (K\cup Z \cup Z')$. Let $C_1\in \Cres$ be induced by $q_1 = |Z|$ and
  $\succ_1$ such
  that $\{x \in X: x\succ_1 \varnothing\} = Z\cup Z'$ and for each $x'\in
  Z'$ and each $x\in Z\setminus Z'$, $x'\succ_1 x$. Then, $C_1(Z\cup
  \{b\}) = Z$ and $C_1(Z\cup Z'\cup\{b\}) = 
  Z'$.
 Let $C_2\in \Cres$ be induced by 
$\succ_2$ such
that  $\{x\in X:x\succ_2 \varnothing\} =  \{a,b\}$ and  $a\succ_2 b
\succ_2 \varnothing$ and $q_2 = 1$. 

Let $C= \mathcal L_E(C_1,C_2)$. Then, $b \in C(Z\cup Z'\cup \{b\})=
Z'\cup \{b\}$ but $b \notin C(Z\cup \{b\})= Z$, which violates path
independence.  
\item[Case 2 (\boldmath  $K\cup Z\cup Z' = X$):]
  Since $Z$ and $Z'$ are finite, there
  is some $b\in X\setminus (Z\cup Z')$. Since $Z\in Dom(R_E)$, there is some $c\in
  X\setminus(Z\cup K) \subseteq Z'$. Since $a\in Z$, $c\neq a$. Let $Z'' = (Z' \setminus \{c\})\cup
  \{b\}$. Since $a\in Z''$, 
  $Z\cup Z'' \cup K= X \setminus \{c\}$ and $|Z''| = |Z'|$. By Case 1,
   $a\in R_E(Z)$ implies that  $a\in R_E(Z'')$.
Since $b\notin Z'$, $Z'\not \subseteq Z''$ and $Z'\cup Z'' \cup
   K = Z'\cup K\subset X$. So, again by Case 1, $a\in R_E(Z'')$
   implies $a\in R_E(Z')$.
\end{description}
Thus, for each $n\in \mathbb N \cup \{0, \infty\}$, letting $T^n =
\cup_{Z\in Dom(R_E), |Z| = n} R_E(Z)$, we have that for each $Z\in
Dom(R_E)$,  $R_E(Z) = Z\cap T^{|N|}$. By monotonicity of $R_E$ on
$Dom(R_E)$ and since $T^0 = \emptyset$, we conclude that $\emptyset =
T^0 \subset T^1\subseteq T^2 \subseteq \dots$. Thus, $R_E$ is
cardinal-linear on $Dom(R_E)$.
\end{proof}

\begin{proof}
  [{\bf\em Proof of \Cref{cl: k disjoint}}]
Suppose $R_E$ is not $K$-disjoint on $Dom(R_E)$. Then, there is 
$Z\in  Dom(R_E)$ with some $a\in R_E(Z)\cap K$.  
Since $Z\in Dom(R_E)$ and $G_E$ is all-or-nothing, $G_E(Z)=Z\cup K
\subset X$. So, there is  $b\in X\setminus(Z\cup K)$. Since
$Z$ is finite, there is some $Z'\subseteq X\setminus (Z\cup\{b\})$
such that $|Z'| = |Z|$. By construction, $ a\notin Z'$. Since $G_E$ is
all-or-nothing, $a\in K \subseteq E(Z')$.
Let $C_1\in \Cres$ be induced by
$\succ_1$ such that $\{x\in X:x\succ_1
\varnothing\} =  Z\cup Z'$ and $x\succ_1 y$ for each $x\in Z'$ and
$y\in Z\setminus Z'$  and $q_1 = |Z'|$. Let $C_2\in \Cres$ be
induced by  $\succ_2$ such that  $\{x\in X:x\succ_2
\varnothing\} =  \{a,b\}$ and  $a\succ_2 b$ and $q_2 = 1$. Let $C=
\mathcal L_E(C_1,C_2)$. Then, $b \in C(Z\cup Z' \cup \{b\}) =
Z'\cup\{b\}$ but $b \notin C(Z \cup \{b\}) =Z$, which violates
path independence. So, $R_E$ is $K$-disjoint on $Dom(R_E)$. 
\end{proof}

\begin{proof}
  [{\bf\em Proof of \Cref{thm: cres}}]We first show necessity and then
  sufficiency.

  \noindent \textbf{Necessity:}
By \Cref{cl: general nec cres} and \Cref{cl: k disjoint}, 
for $\mathcal L_E$ to preserve path independence over $\Cres$, it is
necessary that
\begin{enumerate}
\item $G_E$ is
  threshold-linear.
\item $R_E$ is cardinal-linear and $K$-disjoint on
  $Dom(R_E)$. 
\end{enumerate}
These together imply that $E$ is threshold-linear with cardinal reuse.
This establishes the necessity part of the theorem.

\noindent \textbf{Sufficiency:} We show that if $E$ is
threshold-linear with cardinal reuse, 
then $\mathcal L_E$ preserves path independence over
$\Csubsm$.\footnote{\label{fn:T1 stronger}Note that this is stronger than what \Cref{thm: cres}
  claims, which is sufficiency over only $\Cres$, a subset of
  $\Csubsm$.}

Suppose $E$ is parameterized by   $K\subset X$, $t\in
\mathbb N\cup\{0, \infty\}$, and $\{T^s\}_{s=1}^t$.

Let $C=\mathcal L_{E}(C_1,C_2)$ and fix $Y\subset Y' \in [X]^*$ and
$x\in Y$ such that $x \in C(Y')$. We show that $x\in C(Y)$.  Let $Z = C_1(Y)$ and $Z' =
C_1(Y')$. If $x\in Z$ we are done since $Z\subseteq C(Y)$. If $x\in
Z'$, then since $C_1$ is path independent, $x\in Z$ and we are again
done. It remains to consider the case where $x\notin Z'$ and
$x\notin Z$. Since $x\in C(Y')$, this means $x\in C_2(Y'\setminus
E(Z'))$. Then, since $x\notin  G_E(Z')$,  $G_E(Z') \neq X$
and $|Z'| <t$. So, $E(Z') = (Z'\setminus T^{|Z'|})\cup K$. This
implies that $x\notin K$. Thus, since $x\notin Z$ and $x\notin K$, it
follows that $x\notin E(Z)$ and therefore $x\in Y\setminus E(Z)$.

Since $C_2$ is path independent, it suffices to show that $Y\setminus
E(Z) \subseteq Y'\setminus E(Z')$.
Since $C_1$ is size monotonic and $Y\subset Y'$, $|Z|\leq |Z'| <
t$. Thus, $E(Z) = (Z\setminus T^{|Z|})\cup K$ and $T^{|Z|}\subseteq T^{|Z'|}$.

Since $C_1$ is path independent, $Z'\cap Y \subseteq Z$, so
$(Z'\cap Y)\setminus T^{|Z'|} \subseteq Z\setminus
T^{|Z|}$. Removing both sides from $Y$,
\[
  Y\setminus  (Z\setminus T^{|Z|}) \subseteq     Y\setminus  ((Z'\cap
  Y)\setminus T^{|Z|}) \subseteq Y\setminus  ((Z'\cap
  Y)\setminus T^{|Z|}) \cup [(Y'\setminus Y)\setminus (Z'\setminus
  T^{|Z'|})] = Y'\setminus (Z'\setminus T^{|Z'|}).
\]
Removing $K$ from both sides,
\[
  Y\setminus E(Z) = Y\setminus ((Z\setminus T^{|Z|})\cup K)
  \subseteq
  Y'\setminus ((Z'\setminus T^{|Z'|})\cup K) = Y'\setminus E(Z').
\]
\end{proof}

\subsection{Proofs From \Cref{sec:consistency-csubirc}}
\begin{proof}
  [{\bf\em Proof of \Cref{prop: pi}}]We first show necessity and then
  sufficiency.

  \noindent \textbf{Necessity:} By \Cref{thm: cres}, since
  $\Cpi\supset \Cres$, if $\mathcal L_E$ preserves path independence
  over $\Cpi$, it is threshold-linear with cardinal reuse with respect
  to some $t\in \mathbb N\cup\{0,\infty\}$, $K\subseteq X$ and
  $\{T^n\}_{n=0}^t$.
  
  First, we show that $t$ is  necessarily in $
  \{0,1,\infty\}$. Suppose otherwise, that $1 < t < \infty$. Let
  $Z\subset X$ be such that $|Z| < t$ and $Z\cup K\not\subseteq X$ so
  that $G_E(Z) = Z\cup K \neq X$. Since $Z$ is finite, there is $Z'\subset X$ such that 
  $Z'\not\supset Z$, $|Z'| = t$, and $Z'\cup Z\cup K\neq
  X$. Then, by  threshold-linearity, $G_E(Z) = X$.

  Let $a\in X\setminus(Z\cup Z'\cup K)$ and let $C_1\in \Cpi$ be such
  that $C_1(Z\cup Z' \cup \{a\}) = Z$ and $C_1( Z' \cup \{a\}) =
  Z'$.\footnote{\label{fn:aizerman rationalization} Since $C_1$ can be expressed as the union of
    maximizers of $t$ linear orders over $X\cup \{\varnothing\}$, by
    \cite{AizermanMalishevski:IEEE1981} it is path independent. The
    following is a description of  specific linear orders
    $\{\succ_i\}_{t=1}^t$  that
    rationalize $C_1$ in this way. Let $\{z_1, \dots, z_l\} = Z\cap Z'$,
    $\{z_{l+1},\dots, z_n\} = Z\setminus Z'$, and
    $\{z'_{l+1},\dots, z'_t\} = Z'\setminus Z$. Since $|Z'| = t >
    |Z|$, $t > n$. For each $i$ between 1 and $t$, let $\succ_i$
    be such that if $i
    \leq l$ then $\{x\in X:x\succ_i\varnothing\} = \{z_i\}$, if $l < i
    < n$ then  $\{x\in
    X:x\succ_i\varnothing\} = \{z_i, z'_i\}$ with 
    $z_i \succ_i  z'_i$, and if $i\geq n$ then  $\{x\in
    X:x\succ_i\varnothing\} = \{z_n, z'_i\}$  with
    $z_n \succ_i  z'_i$.}
  Let $C_2 \in \Cres$ be rationalized by $\succ_2$ and $q_2=1$ where
  $\{x\in X:x\succ_2 \varnothing\} = \{a\}$.
  Let $C= \mathcal L_E(C_1,C_2)$. Then, $a \in C(Z\cup Z'\cup \{a\}) =
  Z\cup \{a\}$ but $a \notin C(Z'\cup\{a\}) = Z'$, which violates
  path independence. Therefore, $t\in \{0,1,\infty\}$.

  To complete the proof of necessity, we show that if $t = \infty$
  then, $T^2 = T^3 =\dots$. Suppose, to the contrary, $t= \infty$ but
  for some $l \geq 2$, $T^l \neq T^{l+1}$. Since $T^{l+1} \supset
  T^l$, there is $a\in T^{l+1} \setminus  T^l$. Let $Z\subseteq X$ be
  such that $a\in Z$ and $|Z| = l$. Since $2 \leq |Z| <\infty$ there
  exists $Z'\subseteq X$ such that $a\in Z\cap Z'$, $|Z'| = l+1$,
  $Z\not\subset Z'$, and $Z \cup Z' \cup K \neq X$. Let $b\in
  X\setminus (Z \cup Z' \cup K)$ and let $C_1\in\Cpi$ be such that
  $C_1(Z'\cup\{b\}) = Z'$ and $C_1(Z\cup Z' \cup \{b\}) =
  Z$.\footnote{The argument for why $C_1\in \Cpi$ is identical to that
    in  \Cref{fn:aizerman rationalization}.}
  Let $C_2 \in \Cres$ be rationalized by $\succ_2$ and $q_2=1$ where
  $\{x\in X:x\succ_2 \varnothing\} = \{a,b\}$ and $a\succ_2 b$.
  Let $C= \mathcal L_E(C_1,C_2)$. Then, $b \in C(Z\cup Z'\cup \{b\}) =
  Z\cup \{b\}$ but $b \notin C(Z'\cup\{b\}) = Z'$, which violates
  path independence. Therefore, $T^2 = T^3 = \dots$.

  \noindent \textbf{Sufficiency:} Suppose $E$ is as described in the
  statement of \Cref{prop: pi}.   If $t = 0$, then we are done since
  $C = C_1$ and $C_1$ is path independent.  So, suppose $t\neq 0$.

  Let $Y,Y'\in [X]^*$ be such that $Y\subseteq Y'$. Let $C= \mathcal
  L_E(C_1,C_2)$. Suppose $x\in C(Y')$ is such that $x\in Y$. It
  suffices to show that $x\in C(Y)$.

  First consider $t = 1$.  If
  $x\in C_1(Y')$, then $x\in C_1(Y) \subseteq C(Y)$ since $C_1$
  is path independent. Otherwise, $x\in C_2(Y'\setminus E(C_1(Y')))
  \setminus C_1(Y')$. Since $G_E(Z) = X$ for every nonempty $Z\in
  [X]^*$ and  $x\in Y'\setminus E(C_1(Y'))$, we conclude that $C_1(Y')
  =  \emptyset$ and $x\notin K$.  Since $C_1$ is consistent, $C_1(Y) =
  \emptyset$. Then, since $x\notin K$, $x\in Y\setminus
  E(C_1(Y))$. Moreover, $Y\setminus E(C_1(Y)) = Y\setminus K \subseteq
  Y'\setminus K = Y'\setminus E(C_1(Y'))$. Since $C_2$ is path
  independent and $x\in  C_2(Y'\setminus E(C_1(Y')))$, it then follows 
  that $x\in  C_2(Y\setminus E(C_1(Y))) \subseteq C(Y)$. 

  We complete the proof of sufficiency by considering the case of
  $t=\infty$. Let $Z=  C_1(Y)$ and $Z' = C_1(Y')$. If $x\in Z$, then
  $x\in Z 
  \subseteq C(Y)$ directly. If $x\in Z'$, then again $x\in Z \subseteq
  C(Y)$ since $C_1 \in \Cpi$. So, suppose $x\in C_2(Y'\setminus E(Z'))
  \setminus (Z \cup Z')$.

  Since $C_2\in \Cpi$, it suffices
  to show that $x\in Y\setminus E(Z) \subseteq Y'\setminus E(Z')$.

  Given that $x\in Y'\setminus E(Z')$ and $K\subseteq E(Z')$, we
  conclude that $x\notin K$. Since $E(Z) \subseteq Z\cup K$ and $x\notin
  Z$, it follows that $x\notin E(Z)$. Thus, since $x\in Y$, $x\in
  Y\setminus E(Z)$.

  If $Y\setminus E(Z) \not \subseteq Y'\setminus E(Z')$, there is
  $a\in Y\setminus E(Z)$ such that $a\notin  
  Y'\setminus E(Z')$.  Then, $a\in E(Z')$.
  If $a\in K$, then $a\in
  E(Z)$, contradicting $a\in Y\setminus E(Z)$, so $a\notin K$. Thus,
  since $a\in E(Z') = (Z'\setminus T^{|Z'|})\cup K$ and $a\notin K$,
  $a\in Z'\setminus T^{|Z'|}$.
  
  If  $|Z'| > 1$, then regardless
  of $|Z|,$ $T^{|Z|} \subseteq T^{|Z'|}$. So, since $a\notin
  T^{|Z'|}$, $a\notin T^{|Z|}$. Since $a\notin E(Z)$, we then
  conclude that $a\notin Z$. However, this contradicts the path
  independence of $C_1$ since $a\in Z'$.
  Otherwise, $|Z'| = 1$ and since $a\in Z'\setminus
  T^{|Z'|}$, $Z' = \{a\}$. Since $a\in Y\subseteq Y'$ and $C_1$ is
  path independent, $Z = Z' = \{a\}$. Then, $E(Z) = E(Z')$,
  contradicting 
  $Y\setminus E(Z)\not\subseteq  Y'\setminus E(Z')$. 
\end{proof}

\subsection{Proofs From \Cref{sec:size-monot-csubsm}}
\begin{proof}
  [{\bf\em Proof of \Cref{prop: sm}}] We first show necessity and then
  sufficiency. 

  \noindent \textbf{Necessity:} By \Cref{thm: cres},  if $\mathcal
  L_E$ preserves path independence 
  over $\Cres$, it is threshold-linear with cardinal reuse with respect
  to some $t\in \mathbb N\cup\{0,\infty\}$, $K\subseteq X$ and
  $\{T^n\}_{n=0}^t$.

  Suppose $|X\setminus K| > 1$. We first show that for each $n < t,
  T^n = \emptyset$. Then, we show that $t = \infty$. 

    Suppose, for some $n < t$,  $T^n$  contains 
   $a\in X$. Then there is  $b\in
  X\setminus( K\cup\{a\})$. Let $Z\subseteq X\setminus\{a,b\}$ be such
  that $|Z| = s$. Since $T^n \neq \emptyset$, $n > 0$ and therefore
  there is $c\in Z$. Let $C_1\in \Cres$ be induced 
  by $\succ_1$ such that $\{x\in
  X:x\succ_1 \varnothing\} = Z\cup \{a\}$ and, for each $z \in Z$,
  $a\succ_1 z \succeq_1 c$ and $q_1 = n$.  Let $C_2\in \Cres$ be induced by
  $\succ_2$ such that $\{x\in X:x\succ_2 \varnothing\}
  = \{a,b\}$ and $a\succ_2 b$ and $q_2 = 1$. Let $C= \mathcal L_E(C_1,C_2)$.

  $C_1(Z\cup \{b\}) = Z$. Since $b \notin K$ and $n < t$,  $b\in (Z\cup
  \{b\})\setminus E(Z)$. Since $a\notin Z, C_2(Z\cup
  \{b\})\setminus E(Z)) = \{b\}$. Thus, $C(Z\cup \{b\}) = Z\cup\{b\}$
  so $|C(Z\cup \{b\})|=n+1$.

  $C_1(Z\cup\{a,b\}) = \{a\}\cup (Z\setminus \{c\})$. Since $a\in T^n$
  and $|\{a\}\cup (Z\setminus \{c\})| = n$,  $a\in (Z\cup
  \{a,b\})\setminus E(\{a\}\cup (Z\setminus \{c\}))$, so $C_2((Z\cup
  \{a,b\})\setminus E(\{a\}\cup (Z\setminus \{c\}))) = \{a\}$. Thus, $C(Z\cup
  \{a,b\}) = \{a\}\cup (Z\setminus \{x\})$ so  $|C(Z\cup
  \{a,b\})|=n$. This violates size monotonicity.

  We now show that $t = \infty$. Suppose, for the sake of
  contradiction that  $t < \infty$. Let $a,b\in X\setminus
  K$ and $Z\subseteq X\setminus\{a,b\}$ be such that  $|Z| = t$. Since
  $K\neq X$, $t>0$. Thus, there is 
  $c\in Z$.
  
 Let $C_1\in \Cres$ be induced
  by $\succ_1$ such that $\{x\in
  X:x\succ_1 \varnothing\} = Z$ and $q_1 = t$. Let $C_2\in \Cres$ be induced by
  $\succ_2$ such that $\{x\in X:x\succ_2 \varnothing\}
  = \{a,b\}$ and $q_2 = 2$. Let $C= \mathcal L_E(C_1,C_2)$.

  By definition of $C_1$,  $C_1(Z\cup\{a,b\}) = Z$.
Since $|Z| = t$, $G_E(Z) = X$. Thus, 
 $C(Z\cup\{a,b\}) =   Z$.

  By definition of $C_1$, $C_1((Z\setminus \{c\})\cup\{a,b\}) =
  Z\setminus \{c\}$. Since  $T^{t-1}  = 
  \emptyset$ and $|Z\setminus \{c\}| = t-1$,
  $E(Z\setminus\{c\}) = (Z\setminus\{c\})\cup K$. Thus,
  $((Z\setminus\{c\})\cup\{a,b\})\setminus E(Z\setminus\{c\})) =
  \{a,b\}$ and $C_2(\{a,b\}) =\{a,b\}$. So, $C ((Z\setminus\{c\})
  \cup\{a,b\}) = (Z\setminus\{c\})  \cup\{a,b\}$.

  However, $|C(Z\cup\{a,b\})| =  |Z| = t < t+1 = |(Z\setminus\{c\})
  \cup\{a,b\}| = |C((Z\setminus\{c\})  \cup\{a,b\})| $, in violation
  of size monotonicity.

  \noindent \textbf{Sufficiency:}
  Suppose $E$ is as described in \Cref{prop: sm}. Let $C_1, C_2 \in
  \Csubsm$ and  $C=\mathcal L_{E}(C_1,C_2)$. Fix $Y\subset Y' \in
  [X]^*$. We 
  show that $|C(Y)| \leq |C(Y')|$. Let $Z = C_1(Y)$ and $Z' =
  C_1(Y')$. Since $C_1$ is size monotonic, $|Z| \leq |Z'|$. Since
  $C_1$ is path independent, $Y\setminus Z \subseteq Y\setminus Z'$.
  
  First consider the case of $|X\setminus K| \leq 1$. If $K = X$, then
  $C = C_1$ so we are done. Otherwise, $X\setminus 
  K = \{a\}$. If $a\notin Y$ or if $a\in Z$, then $C(Y) = C_1(Y) =
  Z$. Moreover, $C(Y') 
  \supseteq Z'$, so $|C(Y)| = |Z| \leq |Z'| \leq |C(Y')|$. If $a\in
  Y\setminus Z$, then $a\in Y'\setminus Z'$ so if $a\notin C_2(\{a\})$,
  $|C(Y)| = |Z| \leq |Z'| \leq |C(Y')|$ and if $a\in C_2(\{a\})$,
  $|C(Y)| = |Z|+1 \leq |Z'|+1 \leq |C(Y')|$.

  Now consider $|X\setminus K| > 1$. In this case, $t = \infty$ and
  $T^s = \emptyset$ 
  for each $s$. Thus, $E(\overline Z) = K\cup \overline Z$ for each
  $\overline Z\in [X]^*$ so for each $\overline Y\in
  [X]^*$,  $C_2(\overline Y\setminus E(\overline Z))
  \cap \overline Z = \emptyset$ meaning that
  $|C(\overline Y)| = |C_1(\overline Y)| + C_2(\overline Y\setminus
  E(\overline Z))|$.

  Since $Y\setminus (Z\cup K) \subseteq Y'\setminus (Z'\cup K)$ and
  $C_2$ is size 
  monotonic, $|C_2(Y\setminus (Z\cup K))| \leq |C_2( Y\setminus
  (Z'\cup K))|$. Thus, 
  $|C(Y)| = |Z| + |C_2(Y\setminus (Z\cup K))| \leq |Z'|+ |C_2(
  Y\setminus (Z'\cup K))|  = |C(Y')|$.     
\end{proof}
\subsection{Proofs From \Cref{sec:more-than-two}}

\begin{proof}
  [{\bf\em Proof of \Cref{cl:left-right-comp}}]
We prove these for the case of $m = 3$. However, the proof generalizes
to arbitrary $m\geq 3$.

Letting  $E_1$ and $E_2$ both to be threshold-linear exclusion with
cardinal reuse where $t = N$, $K = \emptyset$, and $T^N = \emptyset$,
right composition   ($\mathcal L_{E_2}(L_{E_1}(C_1, C_2),
C_3)$) implements \Cref{proc:aggregate-quota}. Left
composition ($\mathcal L_{E_1}(C_1, L_{E_2}( C_2, C_3))$) implements
\Cref{proc:individual-quota}.

Next, we show that \Cref{proc:aggregate-quota} cannot be implemented
with left composition. To implement \Cref{proc:aggregate-quota}, $E_1$
is necessarily threshold-linear exclusion with
cardinal reuse where $t = N$, $K = \emptyset$, and $T^N =
\emptyset$. To see this, observe that if we select $C_3\in \mathcal C$ to
be such it always chooses $\emptyset$,  then
\Cref{proc:aggregate-quota} is equivalent to lexicographic composition
with this exclusion function. Setting $C_1\in \mathcal C$ to be such
that it always chooses $\emptyset$, we similarly conclude that  $E_2 =
E_1$. Finally, let $Z_1$ and $Z_2\in [X]^*$ be disjoint such that
$|Z_1| + |Z_2| > N$ but $|Z_1|, |Z_2| < N$. Let $z_3\in X\setminus
(Z_1\cup Z_2)$. Let $C_1, C_2, C_3\in \mathcal C$ be such that
$C_1(Z_1\cup Z_2 \cup \{z_3\}) = Z_1$, $C_2( Z_2 \cup \{z_3\}) = Z_2$,
and $C_3(\{z_3\}) = z_3$. According to \Cref{proc:aggregate-quota},
the final choice ought to be $Z_1\cup Z_2$. However, the left
composition with $E_1$ and $E_2$ yields $Z_1\cup Z_2\cup
\{z_3\}$. Thus,  \Cref{proc:aggregate-quota} cannot be implemented via
left composition.

Finally, we prove that \Cref{proc:individual-quota} cannot be
implemented via right composition. Exactly as argued above, $E_1$ and
$E_2$ are necessarily  threshold-linear exclusion with
cardinal reuse where $t = N$, $K = \emptyset$, and $T^N =
\emptyset$. For the same $C_1, C_2,$ and $C_3$ above, according to
\Cref{proc:individual-quota}, 
the final choice ought to be $Z_1\cup Z_2\cup\{z_3\}$. However, the right
composition with $E_1$ and $E_2$ yields $Z_1\cup Z_2$. Thus,
\Cref{proc:individual-quota} cannot be implemented via 
right composition.
\end{proof}

\subsection{Proofs From \Cref{sec:single-valued-inputs}}
\begin{proof}
  [{\bf\em Proof of \Cref{prop:sv-sub}}]
We first show necessity and then
  sufficiency. 

  \noindent \textbf{Necessity:} We show that 
  if $\mathcal L_E$ preserves path independent  over $\Csvres\times
  \Csvres$, then setting $K=E(\emptyset)$, either
  \begin{enumerate}
  \item for each $x\in X$, $G_E(\{x\}) = X$, or 
  \item there is $T\subseteq X\setminus K$ such that for each $x\in
    X$,
    \[
      E(\{x\}) = \left\{
        \begin{array}{ll}
          K&\text{if } x\in T\\
          K\cup\{x\}& \text{otherwise.}\footnotemark
        \end{array}
        \right.
    \]
  \end{enumerate}
  \footnotetext{This is stronger than the necessity part of
    \Cref{prop:sv-sub} in that the domain $\Csvres\times\Csvres$ is
    smaller than $\Csvres\times\Cpi$, which is the domain for
    \Cref{prop:sv-sub}. }
    First, we show that $K\subseteq G_E(\{a\})$ for each $a\in
  X$. If not, there is $b\in K\setminus E(\{a\})$ such that $b\neq
  a$.   Let $C_1$ be induced by  $\succ_1$ such that $\{x\in
  X:x\succ_1 \varnothing\} = \{a\}$. Let $C_2$ be induced by $\succ_2$ 
  such that $\{x\in
  X:x\succ_1 \varnothing\} = \{b\}$.  Let $C = \mathcal
  L_E(C_1,C_2)$. Then, $C(\{b\}) = \emptyset$ 
  while $C(\{a,b\}) = \{a,b\}$. Since this contradicts path independence,
  $K\subseteq G_E(\{a\})$.

  Second, we show that for each $x\in X$, $G_E(\{x\}) \in \{X,
  \{x\}\cup K\}$. If not, then for some  $a\in X$ there are $b\in
  E(\{a\})\setminus (K\cup \{a\})$ and $c\in X\setminus
  G_E(\{a\}$. By definition, $a, b$, and $c$ are distinct.
  Since $K\subseteq G_E(\{a\})$, $c\notin K$.
  Let $C_1$ be induced by  $\succ_1$ such that $\{x\in
  X:x\succ_1 \varnothing\} = \{a\}$. Let $C_2$ be induced by $\succ_2$ 
  such that $\{x\in
  X:x\succ_1 \varnothing\} = \{b,c\}$ such that $b\succ_2 c$.  Let $C = \mathcal
  L_E(C_1,C_2)$. Then, $C(\{a,b, c\}) = \{a,c\}$ 
  while $C(\{b,c\}) = \{b\}$. Since this contradicts path independence,
  $G_E(\{a\}) = K\cup \{a\}$ or   $G_E(\{a\}) = X$.

Third, we show that if for any pair $a,b\in X$ such that $\{a\}\cup
K, \{b\}\cup K\neq X$, $G_E(\{a\}) = X$ if and only if
$G_E(\{b\}) = X$. Suppose $G_E(\{a\}) \neq X$  but
$G_E(\{b\}) = X$. If there is $c\in X\setminus
(K\cup\{a,b\})$.  
  Let $C_1$ be induced by  $\succ_1$ such that $\{x\in
  X:x\succ_1 \varnothing\} = \{a,b\}$ such that $b\succ_1 a$. Let
  $C_2$ be induced by $\succ_2$  
  such that $\{x\in
  X:x\succ_1 \varnothing\} = \{c\}$.  Let $C = \mathcal
  L_E(C_1,C_2)$. Then, $C(\{a, c\}) = \{a\}$ 
  while $C(\{b,c\}) = \{b,c\}$. Since this contradicts path independence,
  $G_E(\{b\}) = X$. If there is no $c\in X\setminus (K\cup\{a,b\})$,
  then $X = K\cup\{a,b\}$. Let $b'\in K$. Then there is $c\in
  X\setminus (K\cup\{a,b'\})$ and by the above argument, $G_E(\{b'\})
  = X$. Since $K\cup\{b,b'\}\neq X$, we again repeat the argument to
  conclude that $G_E(\{b'\}) = X$ implies $G_E(\{b\}) = X$.

  Fourth, we show that if there is $x\in X$ such that $G_E(\{x\}) \neq
  X$, then for each $x\in K$, $x\in E(\{x\})$. Suppose there
  is $a\in K\setminus E(\{a\})$.
  Since $G_E(\{x\}) \neq X$, $K\neq X$. So, there is
  $b\in X\setminus K$.
  Let $c\in X\setminus \{a,b\}$.  Since $a\in K$, $a\in K\subseteq
  E(\{b\})$. 
  Let $C_1$ be induced by  $\succ_1$ such that $\{x\in
  X:x\succ_1 \varnothing\} = \{a,b\}$ such that $b\succ_1 a$. Let
  $C_2$ be induced by $\succ_2$  
  such that $\{x\in
  X:x\succ_1 \varnothing\} = \{a\}$ such that $a\succ_2 c$.  Let $C = \mathcal
  L_E(C_1,C_2)$. Then, $C(\{a, c\}) = \{a,c\}$ 
  while $C(\{a,b,c\}) = \{b,c\}$. Since this contradicts path independence,
  for each $x\in K, x\in E(\{x\})$.

  To complete the proof of necessity, let $T = \{x\in X: x\notin
  E(\{x\})$. As we have shown above, if there is $x\in X$ such that
  $G_E(\{x\}) \neq X$ then
  $T\cap K = \emptyset.$ By definition, $E(\emptyset) = K$. If
  there is $x\in X$ such that $\{x\}\cap K\neq X$ but $G_E(\{x\}) =
  X$, then by what we have shown above, for every $x\in 
  X$, $G_E(\{x\}) = X$. Otherwise, for each $x\in X,
  G_E(\{x\}) = \{x\} \cup K$.  By definition of $T$, if $x\in T$,
  $x\notin E(\{x\})$ so $E(\{x\}) = K$ and if $x\notin T$,   $x\in
  E(\{x\})$ so $E(\{x\}) = K\cup\{x\}$.

  \noindent \textbf{Sufficiency:}
    Suppose $E\in \mathcal S$ coincides on singletons and the empty
  set  with a threshold-linear
  exclusion function with cardinal reuse where   $K\subset X$, and
    $T\subseteq X\setminus K$.

  Let $C=\mathcal L_{E}(C_1,C_2)$ and fix $Y\subset Y' \in [X]^*$ and
  $x\in Y$ such that $x \in C(Y')$. We show that $x\in C(Y)$.  Let $Z = C_1(Y)$ and $Z' =
  C_1(Y')$. If $x\in Z$ we are done since $Z\subseteq C(Y)$. If $x\in
  Z'$, then since $C_1$ is path independence, $x\in Z$ and we are again
  done. It remains to consider the case where $x\notin Z'$ and
  $x\notin Z$.

If $Y\setminus E(Z)\subseteq Y'\setminus E(Z')$, we are done by
path independence of $C_2$. Otherwise, there is $a\in E(Z')$ such that
$a\notin Z'$ but $a\notin E(Z)$. If $a\notin E(Z)$ then $a\notin
K$. Since $a\in  E(Z')$ and $a\notin K$,  $G_E(Z') = X$
since $a\notin Z'\cup K$. 
Since $G_E(Z') = X$, $C(Y') = Z'$. However, this contradicts
$x\in C(Y')\setminus Z'$.
\end{proof}
\begin{proof}
  [{\bf\em Proof of \Cref{prop:sub-sv}}]
  The proofs of necessity for \Cref{thm: cres} and \Cref{prop: pi}
  appeal only to $C_2\in \Csvres$. Consequently, they establish the
  stronger result that said conditions are necessary even on the
  smaller domains of $\Cres\times\Csvres$ and $\Cpi\times\Csvres$
  respectively. The sufficiency part of \Cref{prop: pi} implies
  sufficiency over the smaller domain as well.  
\end{proof}

\subsection{Proofs From \Cref{sec:equiv}}
\begin{proof} [Proof of \Cref{rmk:many-one}] Sufficiency of this condition is by definition.
For necessity, we proceed by contradiction. Suppose that $Z \in
[X]^*_F$ is such that $x \in Z$ and $y \notin E(Z)$ for $x\sim y$
where $y\neq x$. Let
$C_1\in \Ch$ be such that $C_1(Z \cup \{y\})=Z$.
Let $C_2\in \Cres$ be
induced by $\succ_2$ such that
$\{x\in X:x\succ_2 \varnothing\} = \{y\}$ and $q_2 = 1$. Let $C = \mathcal
L_E(C_1,C_2)$. Then, $C(Z \cup \{y\}) = Z \cup \{y\}$. Since
$x\sim y$, and $x\in Z$, $\mathcal L_E$ does not preserve
many-to-oneness. \end{proof}

\section{Size Monotonicity and Single-valued Choice}
\label{apx:sm-sv}
In this appendix, we consider the preservation of \emph{both} path
independence and size monotonicity when one of the two inputs to the
lexicographic composition is single-valued.
As observed in \Cref{sec:size-monot-csubsm}, compared to
the corresponding results in \Cref{sec:single-valued-inputs},  adding the requirement of
preserving size monotonicity imposes severe restrictions on reuse. 

For the domain $\Csvres\times\Csubsm$, the analog of \Cref{prop:sv-sub}
is as follows. 
\begin{proposition}
  \label{prop:sv-subsm}
  $\mathcal L_E$ preserves path independence and size monotonicity over $\Csvres\times 
  \Csubsm$ if and only if  on the set of singletons and the empty set,
  $E$ coincides with an exclusion function that is threshold-linear
  with cardinal reuse where if $|X\setminus K|>1$, $T^n = \emptyset$
  for all $n$.
\end{proposition}
\begin{proof}
  We first show necessity and then
  sufficiency. 

  \noindent \textbf{Necessity:} From the proof of the necessity part
  of  \Cref{prop:sv-sub}, since $\Csvres \subset \Csubsm$ and 
  $\mathcal L_E$  preserves path independence over
  $\Csvres\times\Csubsm$, 
  either for each $x\in X$, $G_E(\{x\}) = X$ or there is $T\subseteq
  X\setminus K$ such that for each $x\in X$, $G_E(\{x\}) = \{x\}\cup
  K$ and $R_E(\{x\}) = \{x\}\cap T$. In the former case, we complete
  the proof by noting that  $K = X$.  Suppose, instead that the latter
  is the case. If, $|X\setminus K| \leq 1$, we are again done. So,
  supposed  $|X\setminus K| > 1$. We show that $T =
  \emptyset$. Suppose there   is $ a\in T$. Since $|X\setminus K| > 1$, 
  there is $b\in X\setminus (K\cup\{a\})$. Let $c\in X\setminus\{a,b\}$.
  Let $C_1$ be induced by  $\succ_1$ such that $\{x\in
  X:x\succ_1 \varnothing\} = \{a,c\}$ and $a\succ_1 c$. Let $C_2$ be
  induced by $\succ_2$  
  such that $\{x\in
  X:x\succ_1 \varnothing\} = \{a,b\}$ and $a\succ_2 b$.  Let $C =
  \mathcal L_E(C_1,C_2)$. Then, $C(\{a,b, c\}) = \{a\}$ while
  $C(\{b,c\}) = \{b,c\}$. Since this contradicts size monotonicity,
  $T=\emptyset$. 

  \noindent \textbf{Sufficiency:}  This follows from the sufficiency
  part of \Cref{prop: sm}.  
\end{proof}

For the domain $\Csubsm \times \Csvres$, the analog of \Cref{prop:sub-sv}
is as follows. 
\begin{proposition}
  \label{prop:subsm-sv}
    $\mathcal L_E$ preserves path independence and size monotonicity  over $\Csubsm\times \Csvres$ if and only
    if $E$ is threshold-linear with cardinal reuse with respect to
    $K\subset X$, $t\in \mathbb N\cup\{0,\infty\}$, and
    $\{T^n\}_{n=0}^t$ such that for each $l$, $T^l =  \emptyset$ if
    $|X\setminus K|>1$.
\end{proposition}
\begin{proof}
  We first show necessity and then
  sufficiency. 

  \noindent \textbf{Necessity:} 
  Since the proof of necessity for \Cref{thm: cres} only appeals to
  $C_2\in \Csvres$, it follows that for $\mathcal L_E$ to preserve
  path independence over $\Csubsm\times\Csvres$, $E$ is
  threshold-linear with cardinal reuse. The part of the proof of
  \Cref{prop: sm} where we establish that $|X\setminus K| > 1$ implies
  that for each $n, T^n = \emptyset$ also relies only on $C_2\in
  \Csvres$. Together, these observations complete the proof of necessity.

  \noindent \textbf{Sufficiency:}  As noted in \Cref{fn:T1 stronger},
  the sufficiency part of \Cref{thm: cres} implies sufficiency for
  preserving path independence.  So, it remains to show that for each
  pair 
$(C_1,C_2)\in  \Csubsm \times ~\Csvres$, if $E$ is  threshold-linear
with cardinal reuse such that either $|X\setminus K| \leq 1$ or for
all $n, T^n = \emptyset$, 
$C = \mathcal L_E(C_1,C_2)$ is size monotonic.

If $|X\setminus K|\leq 1$, then the proof is  exactly as
in the proof of \Cref{prop: sm}. So, suppose $|X\setminus K| >
1$. Then,  for each $n, T^n = \emptyset$ so that for each $Z\in [X]^*, 
E(Z) = Z\cup K$. 
Fix $Y\subset Y' \in [X]^*$. Let $C_1(Y') = Z'$ and $C_1(Y) =
Z$. There are two cases to consider. 

  \begin{description}

  \item[Case 1 {\boldmath($|Z|<|Z'|$)}:]
    Since $C_2$ is single-valued, regardless of $E(Z)$ and $E(Z')$, we have $|C_2(Y\setminus E(Z))| - |C_2(Y'\setminus E(Z'))| \leq 1$. Thus, $C(Y) = |Z| + |C_2(Y\setminus E(Z))| \leq |Z'| +|C_2(Y'\setminus E(Z'))| = C(Y')$.

\item[Case 2 {\boldmath($|Z|=|Z'|$)}:]  Since $C_1$ is path
  independent, $Y\setminus Z \subseteq Y'\setminus Z'$. Removing
  $Z\cup K$ from both sides gives $Y\setminus (Z\cup K) \subseteq Y'
  \setminus (Z'\cup Z\cup  K) \subseteq Y'\setminus (Z'\cup K)$. There
  are two subcases to consider. 

(a) If $|Z| = |Z'| < t$, then $E(Z) = Z\cup K$ and $E(Z') = Z'\cup K$. Since $C_2$ is size monotonic, $|C_2(Y\setminus E(Z))| = |C_2(Y\setminus (Z\cup K))| \leq |C_2(Y'\setminus (Z'\cup K))| = |C_2(Y'\setminus E(Z'))|$. So, $|C(Y)| = |Z| + |C_2(Y\setminus E(Z))| \leq |Z'| + |C_2(Y'\setminus E(Z'))| = |C(Y')|$.

(b) If $|Z| = |Z'| \geq t$, then $E(Z) = E(Z') = X$, so $|C_2(Y\setminus E(Z))| = |C_2(Y'\setminus E(Z'))| = 0$ and  $|C(Y)| = |Z| = |Z'| = |C(Y')|$.
\end{description}
These two cases complete the proof.
\end{proof}

In contrast with \Cref{prop:sub-sv}, while the added requirement of
preserving size monotonicity precludes most forms of reuse, the input
being size monotonic permits greater flexibility in the threshold.

\section{Necessity of Weak All-Or-Nothingness}
\label{apx:weak-AorN}
In \Cref{ex:all or nothing not necessary} the exclusion function  jumps
from $K= E(\emptyset)$ to a set $\tilde X \subset X$. This example is
special in that the items in $X \setminus \tilde X$  are all in the
same equivalence class. Our next result shows that this feature is
necessary. A dilation $D \in \mathcal D$ is \textbf{weakly
  all-or-nothing} if, for all $Z \in [X]^*_F$, $D(Z) \in
\{I_Z\cup K,X\}$ if there are $x,y\in X\setminus (I_Z \cup
K)$ such that $x\nsim y$ and otherwise,  $I_Z\cup K
\subseteq D(Z) \subseteq X$.

Before we show necessity of weak all-or-nothingness for the
preservation of path independent completability, we introduce a
necessary condition akin to monotonicity. We say that the dilation  $D
\in \mathcal D$ is \textbf{many-to-one 
  monotonic} if for each $Z, Z'\in [X]_F^*$ such that $Z\subseteq Z',
D(Z) \subseteq D(Z')$. As with our earlier use of the ``many-to-one''
modifier, it means that the condition applies only on $[X]_F^*$ as
opposed to on all of $[X]^*$.
Analogous to the necessity of monotonicity for the preservation of
path independence, we have the following result on preserving path
independent completability.

\begin{claim}
  \label{cl:doctor necess monot}
If $\mathcal L_E$ preserves many-to-oneness and path independent
completability over $\Chres$, then $G_E$ is \textbf{many-to-one monotonic}.  
\end{claim}
\begin{proof}
  By \Cref{rmk:many-one}, $E$ is equivalence-excluding. So, if  $G_E$
  is not many-to-one monotonic then there are $Z,Z'\in 
 [X]^*_F$ such that $Z\subset Z'$ and  $a\in (I_Z\cup
 E(Z))\setminus ( I_{Z'}\cup E(Z'))$.

Let $C_1\in \Chres$
be induced by  $\succ_1$ such that
$\{x\in X:x\succ_1 \varnothing\} = Z'$ and  $q_1=
|Z'|$.  Let $C_2\in \Chres$ be
induced by $\succ_2$ such that
$\{x\in X:x\succ_2 \varnothing\} = \{a\}$ and $q_2= 1$.

Since $a\notin Z'$, $C_1(Z'\cup\{a\}) = Z'$. Since $a\notin I_{Z'}\cup
E(Z')$, $a\in (Z'\cup\{a\})\setminus E(Z')$. Thus,
$C_2((Z'\cup\{a\})\setminus E(Z')) = \{a\}$ and therefore
$C(Z'\cup\{a\}) = Z'\cup \{a\}\in [X]^*_F$.

Since $Z\subset Z'$, $C_1(Z\cup\{a\}) = Z$. Since $a\in E(Z)$,
$a\notin (Z\cup\{a\})\setminus E(Z)$. Thus, 
$C_2((Z\cup\{a\})\setminus E(Z)) = \emptyset$ and therefore
$a\notin C(Z\cup\{a\}) = Z$. This violates path independent
completability of $C$.
\end{proof}

Finally, we show that weak all-or-nothingness is necessary.

\begin{claim}
  \label{cl:doctor necess near aon}
If $\mathcal L_E$ preserves many-to-oneness and path independent
completability over $\Chres$, then $G_E$ is weakly
all-or-nothing.  
\end{claim}
\begin{proof}
  By  \Cref{rmk:many-one} and  \Cref{cl:doctor necess monot}, if
  $\mathcal L_E$ preserves 
  path independent completability, it is  equivalence-excluding and
  $G_E$ is many-to-one monotonic.  Suppose there is $Z\in [X]^*_F$
  such that $X\supset G_E(\underline Z) \supset I_Z\cup K$. Let
  $\underline Z$ be such an element of $[X]^*_F$ such that for all
  $Z\subset \underline Z,  G_ E(Z) =I_Z\cup K$.
  By definition of $\underline Z$, there are $a\in
  X\setminus G_E(\underline Z)$, $b\in \underline Z$, and $c\in G_E(\underline
  Z)\setminus(I_{\underline Z}\cup K)$. Since $c\notin
  I_{\underline Z}\cup K$, $c\notin \underline Z$ so $c\in
  E(\underline Z)$. Note that
  since $a,c\notin I_{\underline Z}$, $(I_a \cup I_c) \cap I_{\underline
    Z} = \emptyset$.

If $a\nsim c$, then $\mathcal L_E$ does not preserve path independent
completability by the same argument given to show necessity of
all-or-nothingness for \Cref{cl: general nec cres}. Thus, $a\sim
c$. Since this applies 
for every $a\in X\setminus G_E(\underline Z)$ and
$c\in G_E(\underline
Z)\setminus (I_{\underline Z}\cup K)$, there is some item
$d\in X$
such that $X\setminus (I_{\underline Z}\cup E(\emptyset)) =
(X\setminus G_E(\underline Z))\cup (G_E(\underline Z)\setminus 
(I_{\underline Z}\cup K)) \subseteq I_d$. So, for each pair
$x, y\in X\setminus (I_{\underline Z}\cup K)$, $x\sim y$.

To complete the proof, we observe that for any $Z\in [X]^*_F$ such
that $Z\supset \underline
Z$, since $I_{Z} \supseteq I_{\underline Z}$, 
$X\setminus (I_{Z}\cup K)
\subseteq X\setminus (I_{\underline Z}\cup K)$. So, there are items
from at most one equivalence class in $X\setminus (I_{Z}\cup K)$.
\end{proof}

In our base model we assumed that $X$ is countably infinite in order
to avoid issues that arise at the boundaries---that is, for sets of
size $|X|-1$. While this issue looks similar to the issue with having
to weaken  all-or-nothingness, 
requiring the number of equivalence classes to be countably infinite
does not change it. The 
weakening of all-or-nothingness is driven by the finiteness
of equivalence classes in $X\setminus K$. Since $K$ may be
countably infinite, even if $X$ is countably infinite, $X\setminus
K$ may still be  finite.

\bibliography{References}

\end{document}